\documentclass[conference,compsoc]{IEEEtran}

\ifCLASSOPTIONcompsoc
  \usepackage[nocompress]{cite}
\else
  \usepackage{cite}
\fi

\usepackage[pdftex]{graphicx}
\usepackage{booktabs}
\usepackage{multirow}
\usepackage{makecell}
\usepackage{colortbl}
\usepackage{threeparttable}
\usepackage{array}
\ifCLASSOPTIONcompsoc
  \usepackage[caption=false,font=footnotesize,labelfont=sf,textfont=sf]{subfig}
\else
  \usepackage[caption=false,font=footnotesize]{subfig}
\fi

\usepackage{amsmath}
\usepackage{amssymb}
\usepackage{algorithm}
\usepackage[noEnd=true,indLines=true,commentColor=sysgray]{algpseudocodex} %
\usepackage{tabularx} %

\usepackage{siunitx}
\usepackage{pifont} %

\usepackage{xcolor}
\usepackage{tikz}
\usetikzlibrary{arrows.meta,positioning,calc,fit,backgrounds,shapes.geometric,shapes.multipart,decorations.pathreplacing}
\usepackage{pgfplots}
\pgfplotsset{compat=1.18}
\usepgfplotslibrary{groupplots}

\usepackage{url}
\usepackage[hidelinks,breaklinks=true]{hyperref}
\usepackage[capitalise,noabbrev]{cleveref}

\definecolor{tabhdr}{HTML}{EDF2F7}
\definecolor{tabalt}{HTML}{EAF0F7}
\definecolor{sysblue}{HTML}{2C5AA0}
\definecolor{sysgreen}{HTML}{2E7D4F}
\definecolor{sysorange}{HTML}{C2772A}
\definecolor{sysgray}{HTML}{4A5568}
\definecolor{oiblue}{HTML}{0072B2}
\definecolor{oiorange}{HTML}{E69F00}
\definecolor{oigreen}{HTML}{009E73}
\definecolor{oivermil}{HTML}{D55E00}
\definecolor{oipurple}{HTML}{CC79A7}
\definecolor{oisky}{HTML}{56B4E9}
\definecolor{intD1}{HTML}{1F5FA8}\definecolor{intD2}{HTML}{11806A}\definecolor{intD3}{HTML}{7B4FA0}
\definecolor{intD4}{HTML}{C8581A}\definecolor{intD5}{HTML}{8C6A12}\definecolor{intD6}{HTML}{B0304E}

\makeatletter
\renewcommand\paragraph{\@startsection{paragraph}{4}{\z@}%
  {1.0ex plus .4ex minus .2ex}{-1sp}{\normalfont\normalsize\bfseries}}
\makeatother

\newcommand{\sys}{\textsc{AutoPI}}
\newcommand{\tax}{\textsc{PIAtlas}}
\newcommand{\corp}{\textsc{PICorpus}}
\newcommand{\sysL}{\textsc{AutoPI}\textsubscript{L1}}
\newcommand{\sysF}{\textsc{AutoPI}\textsubscript{Full}}
\newcommand{\sysSE}{\mbox{\textsc{AutoPI}\textsubscript{\smash[b]{Seed-Exploit}}}}

\usepackage{fontawesome5}
\begin{document}

\title{\sys{}: Taxonomy-Guided Adaptive Red Teaming\\ for Prompt Injection in LLM Agents}

\newsavebox{\arxivlinkbox}
\sbox{\arxivlinkbox}{\normalfont\normalsize\color{sysblue}%
\href{https://stevendamianakis.github.io/autopi-site/}{\faGlobe\enspace stevendamianakis.github.io/autopi-site}%
\hspace{2.2em}%
\href{https://github.com/stevendamianakis/autopi}{\faGithub\enspace github.com/stevendamianakis/autopi}}
\newlength{\arxivauthordepth}
\newcommand{\arxivlinkline}{%
\setlength{\arxivauthordepth}{\prevdepth}%
\nointerlineskip\vskip-\arxivauthordepth
\vskip\dimexpr 44pt-\ht\arxivlinkbox\relax
\hbox to\linewidth{\hfil\usebox{\arxivlinkbox}\hfil}%
\nointerlineskip\vskip-\dimexpr 44pt+\dp\arxivlinkbox-\arxivauthordepth\relax
\hbox{}}
\long\def\arxivlinksafterauthors#1\par{#1\par\arxivlinkline}
\author{\IEEEauthorblockN{Stavros Damianakis}
\IEEEauthorblockA{University of Southern California\\
sdamiana@usc.edu}
\and
\IEEEauthorblockN{Konstantinos Psounis}
\IEEEauthorblockA{University of Southern California\\
kpsounis@usc.edu}\arxivlinksafterauthors}
\renewcommand{\IEEEtitletopspaceextra}{-2\baselineskip}

\IEEEtitleabstractindextext{%
\begin{abstract}
Prompt injection is the dominant unsolved security risk for LLM agents, and the
community now agrees that robustness must be measured against adaptive attackers
rather than fixed benchmark injections. Many adaptive red teamers exist, and most
are built to maximize attack success rate (ASR). Yet ASR counts how often a
target breaks, not in how many different ways. A red teamer can achieve a high
ASR by reusing a few techniques that work, leaving the others untested and
absent from the attacks used to harden models. We present \sys{}, an adaptive
red teamer designed to find many different ways to break a target, not just to
break it often. \sys{} uses \tax{}, a taxonomy of prompt-injection techniques,
both to guide its search and to measure coverage. For each scenario, an attacker
LLM writes an injection for every sub-technique in the taxonomy and then reworks
the ones that fail using feedback from an LLM judge. The method is black-box and
training-free, it plugs into existing benchmarks in place of their fixed attack
templates, and it stays up to date as the taxonomy evolves. We evaluate
\sys{} on AgentDojo and IPI Arena against open-weight and proprietary target
models. Alongside ASR, we report \emph{technique coverage}, the fraction of
taxonomy classes, techniques, and sub-techniques that yield at least one
successful injection. On AgentDojo, \sys{} finds a successful injection against
Qwen3.6-35B in all six classes of the taxonomy and for more than 90\% of its
techniques. Against Claude Haiku~5.5, the most robust target we evaluate, it
achieves between 20\% and 50\% higher ASR and between 25\% and 400\% higher
class-level coverage than recent state-of-the-art adaptive red teamers, while
requiring no training. We release our code and the generated attacks to support
reproducible, coverage-aware evaluation.
\end{abstract}

}

\maketitle
\IEEEdisplaynontitleabstractindextext

\section{Introduction}
\label{sec:intro}

Agents built on large language models (LLMs) now read email, write and run code,
browse the web, and operate desktop computers for their users. To do so they
read content that neither the user nor the developer wrote, such as
retrieved documents, tool results, emails, and web pages.
\emph{Prompt injection} hides instructions in that content. The model follows the
planted text as a command instead of processing it as data, and the agent carries
out an action the attacker chose~\cite{greshake2023not,liu2024formalizing,zhan2024injecagent}.
Prompt injection has held first place in the OWASP Top~10 for LLM applications
since its first edition in
2023~\cite{owasp2023llmtop10,owasp2024llmtop10,owasp2026llmtop10}, models still
fail to keep instructions apart from data~\cite{zverev2025can,owasp2026llmtop10},
and a growing view holds that the vulnerability can be contained but is unlikely
ever to be eliminated~\cite{willison2025lethal,openai2025continuously,abdelnabi2026alwaysfall,owasp2026llmtop10}.

\input{sections/fig_intuition}

\paragraph*{Robustness is an adaptive-attack property}
Because injection is unlikely to be eliminated, vendors and researchers measure
\emph{how hard} it is to achieve. The central methodological lesson of the past
two years is that this measurement must use attackers that \emph{adapt} to the
target. Defenses that look strong against fixed benchmark injections collapse under resourced adaptive attackers. Most of twelve jailbreak and injection defenses, the majority of
which originally reported near-zero attack success rate (ASR), exceed 90\% ASR once the attacker adapts~\cite{nasr2026attacker}, and white-box adaptive
attacks push all eight agent defenses tested in~\cite{zhan2025adaptive} above
50\%. Attackers that rewrite injections against a given agent likewise reach
substantially higher ASR than static
attacks~\cite{shi2025lessons,ma2026autodojo,wang2025agentvigil}. Frontier vendors agree. A recent system card states, ``Fixed datasets of known attacks can provide a
false sense of security, as a model may perform well against established attack
patterns while remaining vulnerable to novel approaches''~\cite{anthropic2026opus55card}. As targets harden, a static
benchmark increasingly measures the staleness of its own templates rather than
the robustness of the model.

\paragraph*{Coverage matters}
Adaptivity shows how hard a target is to break. Coverage shows which ways of
breaking it were tested and which still work. Robustness is a worst-case
property, so an evaluation that omits an injection technique (a way of constructing the injected text) can overstate robustness
however adaptive it is within the techniques it exercises. Frontier practice
reflects this. Gemini~4's red-teaming system is rewarded for finding new
manipulation techniques rather than new
wordings~\cite{google2026gemini4}. ASPIRE reports coverage beside ASR
because the two can disagree. An exploit-only search beats
an explore-only search on ASR (45.5\% vs.\ 40.0\%) yet finds fewer than a third
as many unique vulnerabilities, its unit of coverage (24 vs.\ 81)~\cite{google2026aspire}. Good
evaluations and benchmarks must therefore account for coverage, and good red
teamers must produce the diverse attacks that adversarial training and defenses
need (\cref{sec:bg:coverage}).

\paragraph*{What a red-teaming method should optimize}
Red teaming does not exist only to produce a headline ASR. Its successful injections are reused as adversarial-training data that hardens the next model and evaluation data that stresses the next defense~\cite{shi2025lessons,google2026gemini4,wallace2026gptred,wang2026pimine}.
For that downstream purpose, three properties matter.
\textbf{(i) Effectiveness} is how often an attacker succeeds, reported as scenario-level ASR.
\textbf{(ii) Coverage (technique diversity)} is \emph{how many distinct ways} of building
an injection the attacker can make work. An attacker that reuses one trick
across every scenario can post a high ASR yet contribute a thin, repetitive
training set that teaches a defender to recognize one
pattern~\cite{he2026reta,google2026gemini4}.
\textbf{(iii) Efficiency} is the attempts, latency, and cost to get there.
Most automated red teamers optimize (i) and (iii). Reinforcement-learned
attackers are rewarded per successful
attempt~\cite{wen2025rl,yin2026pismith,chen2026learning,anon2026climbing}, and
search- or strategy-based attackers maximize success under a query
budget~\cite{wang2025agentvigil,wang2026pimine}. Public attackers have largely
left property~(ii) implicit. Those that measure diversity mostly use embedding or
token distances~\cite{beutel2024diverse,wen2025rl}, which surface edits can
game~\cite{wen2025rl}. Frontier systems judge diversity
by inspection~\cite{wallace2026gptred} or draw strategies
from unpublished or partly target-specific
sets~\cite{google2026gemini4,google2026aspire}, so
coverage cannot be compared across methods. The field still lacks a shared
yardstick for what ``a different technique'' means.

\paragraph*{Our approach}
We present \sys{}, a taxonomy-guided red-teaming method that turns a
prompt-injection benchmark with fixed injection templates into an adaptive
attack, and that is designed to find substantially different attacks, not copies
of the same attack. \sys{} is a \emph{black-box} attacker. It only sends inputs
and observes the transcript and outcome, so it runs against closed, hosted
models (\cref{sec:threat}). \sys{} is
driven by \tax{}, a recently proposed single-axis taxonomy that organizes
injections by \emph{how the injected instruction is constructed to subvert the
model}. Its six classes refine into techniques and
sub-techniques~\cite{damianakis2026piatlas}. \sys{} attacks each benchmark data
point with three phases that a researcher can combine in any order
(\cref{sec:method}). \emph{Initialization} tries an optional fixed seed injection. \emph{Exploration}
asks an attacker (``generator'') LLM to realize a concrete injection for each
sub-technique of the taxonomy. \emph{Exploitation} reworks failed injections, the seed included, into variants of the same construction under
accumulating feedback from an LLM judge. \cref{fig:intuition} sketches the
intuition. Working injections are rare, and \tax{} conjectures that they
lie in regions, one per technique~\cite{damianakis2026piatlas}, so
exploration takes the search near every region and exploitation closes in on
each. Where prior attackers search their own
strategies, \sys{} sweeps every sub-technique of a
published taxonomy, and coverage counts techniques, not variants. The engine is benchmark- and model-agnostic. A small adapter connects it
to a new benchmark, and the target, the generator, and the judge are
plain LLM endpoints. The generator is \emph{scenario-aware}. It is given the
user's task and the injection goal and writes each injection line directly for
that scenario, a deliberately permissive threat model in the sense a
frontier vendor uses for its adaptive
evaluations~\cite{anthropic2026opus55card}.

\paragraph*{Contributions}
This paper makes five contributions.
\textbf{(1)~A training-free red teamer built for coverage, effectiveness, and efficiency.} \sys{} turns the \tax{}
construction taxonomy into a concrete explore/exploit red-teaming procedure
(\cref{sec:method}). Prior attackers search for what succeeds most often and
tend to reuse a few techniques that work. \sys{} instead sweeps every
sub-technique of the taxonomy and then refines the candidates that fail, so its
successful injections spread over the technique space by design. It is
black-box and needs no training, whereas two of the three recent adaptive
attackers we compare with are trained before the attack.
\textbf{(2)~Coverage as a comparable metric.} We show why coverage is part of
what a robustness evaluation measures (\cref{sec:bg:coverage}) and define
\emph{technique coverage}, the fraction of taxonomy classes, techniques, and
sub-techniques with at least one successful injection. We
report it as a first-class metric beside scenario-level ASR,
with the injections of every attacker labeled by one procedure
(\cref{sec:setup:metrics}).
\textbf{(3)~A steerable trade between ASR and coverage.} We find, as
\cref{fig:intuition} suggests, that exploiting a strong seed wins in fewer
attempts but with fewer techniques, a taxonomy sweep covers far more, and a
weak seed loses on both. The strategy and the taxonomy are inputs, so a
researcher steers this trade (\cref{sec:results:tradeoff}).
\textbf{(4)~A comparison with three recent adaptive attackers.} We evaluate
\sys{} on AgentDojo~\cite{debenedetti2024agentdojo} and
IPI Arena~\cite{dziemian2026how}, against Qwen3.6-35B on both and Claude Haiku~5.5 on
AgentDojo, and compare it with static benchmark seeds,
PIMiner~\cite{wang2026pimine}, CtH~\cite{anon2026climbing}, and
AutoDojo~\cite{ma2026autodojo} (\cref{sec:results}, \cref{fig:scorecard}).
\emph{Coverage.} On AgentDojo against Qwen3.6-35B, \sysF{} covers over 90\% of
\tax{} techniques, the widest coverage of any attacker we run, and no baseline
exceeds 65\%.
\emph{Effectiveness.} \sys{} solves more than 70\% of the scenarios on AgentDojo.
PIMiner and CtH are trained before the attack and \sys{} is not. Even so, its
seed-exploit strategy has the second-highest ASR at equal cost on AgentDojo
against Qwen3.6-35B, above PIMiner and behind only CtH, and \sysF{} comes within
two scenarios of both on IPI Arena, where it runs unchanged.
\emph{Efficiency.} With training counted, one attempt of \sys{} costs 0.6
GPU-seconds, against 2.9 to 14.5 for the other three.
\textbf{(5)~Artifacts.} We release the \sys{} engine, adapters, and the
generated attack corpus (\cref{app:config}) to support reproducible,
coverage-aware red-teaming.

\section{Background}
\label{sec:bg}

\subsection{Prompt injection and adaptive evaluation}
\label{sec:bg:pi}
\label{sec:bg:adaptive}
Adapting earlier formulations~\cite{liu2024formalizing,greshake2023not,damianakis2026piatlas}, a
target model receives a context $C=\langle s,u,d\rangle$ of a system prompt $s$,
a user message $u$, and data segments $d$ (retrieved documents, tool outputs,
web pages, sub-agent messages, memory). The aligned model should follow the
instructions in $s$ and $u$ and treat $d$ as inert data. An \emph{indirect}
prompt injection, our focus, embeds attacker-authored text $x$ into some
$d_i\in d$ so that the model treats $x$ as an instruction and produces a
response that violates the intent of
$s$ or $u$. A benchmark fixes a \emph{data point}, usually a pair
$(\text{user task},\text{injection task})$ (IPI Arena instead ships self-contained scenarios). The user task defines the benign
workflow, and the injection task defines the attacker goal
and a predicate that decides whether it was achieved.
In the agent setting this predicate is usually a program that reads \emph{environment state}~\cite{debenedetti2024agentdojo} or the parsed tool
calls~\cite{zhan2024injecagent}, and some benchmarks add an LLM judge for
criteria a program cannot check~\cite{dziemian2026how}.

Early benchmarks shipped a small set of hand-written injection strings, and
most reported ASRs are still measured against such a
fixed set~\cite{liu2024formalizing,zhan2024injecagent,debenedetti2024agentdojo}.
A fixed set measures robustness to a frozen distribution, which an adaptive
adversary does not respect, and adaptive stress tests have repeatedly
overturned such numbers (\cref{sec:intro}). The resulting prescription is to
evaluate against a resourced attacker that tries as hard as possible to break the
target~\cite{nasr2026attacker} and to state the attempt
budget behind each number~\cite{anthropic2026opus55card,google2026gemini4}.
\sys{} is such an attacker, and we report scenario-level ASR as a function of
budget.

\subsection{Why technique coverage matters}
\label{sec:bg:coverage}

\paragraph*{Coverage and robustness}
An \emph{injection technique} is a way of constructing the injected text. To be
robust, a target must resist every injection technique, because one technique that
works is enough to break it. Let $\mathcal{C}$ be a set of classes that group
related techniques, and let $R_c(f)$ be the survival of target $f$ against a
best-effort attacker restricted to class $c$. An evaluation that
exercises only the classes $S\subset\mathcal{C}$ measures
$\min_{c\in S}R_c(f)\ge\min_{c\in\mathcal{C}}R_c(f)$, which overstates robustness
unless $S$ contains the weakest class. The gap is large in practice. On the same
tool-use scenarios, Gemini~3.8 Flash survives an adaptive attacker in 60\% of
scenarios at 150 attempts, but in only 36\% when the attacker is restricted to
turn injection, a form of structural spoofing~\cite{google2026gemini4}, suggesting
that general adaptive search does not exhaust an individual technique.
Recasting one data
leak likewise lifts its success on GPT-4o from 0\% to as much as
100\%~\cite{rahman2026framing}.
This worst-class survival is an illustration. As a practical metric, we report
the fraction of taxonomy units with at least one successful injection
(\cref{sec:setup:metrics}).

\paragraph*{Coverage and evaluation}
A benchmark's ASR averages over the injection techniques it contains, so it cannot
register a weakness, or a fix, in a technique it omits. Static benchmarks also
saturate quickly~\cite{google2026gemini4}, defenses that succeed on one benchmark
often fail on another, a risk PIArena ties to narrow attack
diversity~\cite{geng2026piarena}, and phrasing alone has a large effect on
AgentDojo's ASR~\cite{debenedetti2024agentdojo}. Evaluations should therefore
report coverage beside ASR. ASPIRE does so, counting a
consequence--method--environment triple as covered once a test succeeds, the same
rule we use~\cite{google2026aspire}. Its results show what ASR hides. MELON,
PI-Guard, and Spotlighting remove 18.8\%, 12.5\%, and 7.3\% of the unique
vulnerabilities, although PI-Guard lowers ASR most, and none reduces the number of working injection methods.

\paragraph*{Coverage and red teaming}
Successful injections become training data, so a red teamer's output determines
what a defender can learn. A search rewarded only for success keeps returning to what already worked~\cite{google2026gemini4}. Two of PIMiner's seven strategies
supply about 80\% of its in-context examples~\cite{wang2026pimine}, RL-Hammer
tends to settle on one pattern~\cite{wen2025rl}, and OpenAI's GPT-Red collapsed onto a narrow class of strategies when
trained against one defender~\cite{wallace2026gptred}.

That concentration reaches the defender. In prompt injection, adding RL-generated attacks that clustered on two
templates to adversarial training raised ASR on an unseen attack family from
25.7\% to 30.7\%, and the authors
conclude that adversarial training must also cover distinct
strategies~\cite{he2026reta}. For jailbreaks, robustness to unseen attacks improves
with skill coverage at fixed training size~\cite{dabas2026dejavu}, and
adversarial training protects least against attacks unlike the one it
trained on~\cite{mazeika2024harmbench}.
The vendors that train frontier models report the same lesson from practice.
Google DeepMind trains Gemini on a broad variety of attacks because models
trained on hand-written attacks may not generalize to unseen
ones~\cite{shi2025lessons}, and it rewards its red teamer for
fundamentally different manipulation techniques and not for new
wordings~\cite{google2026gemini4}. OpenAI requires that its red teamer not
collapse to a single style of attack~\cite{wallace2026gptred}. Anthropic
documents how a gap in training
becomes a gap in the model. Training to recognize injections in tool results
left early snapshots more likely to follow instructions planted in text that a
user pastes~\cite{anthropic2026opus55card}.
Narrow training can pass the tests of its time. SecAlign fine-tunes on two simple
attack types and reports robustness to stronger, unseen
attacks~\cite{chen2025secalign}, yet adaptive attackers
later broke its successor
Meta-SecAlign~\cite{nasr2026attacker,chen2025metasecalign}.

Each of these studies measures diversity over its own strategy set. \sys{} uses \tax{} as a shared partition against
which attackers can be compared (\cref{sec:bg:piatlas}).

\subsection{The \tax{} construction taxonomy}
\label{sec:bg:piatlas}
\sys{} searches over \tax{}, a single-axis taxonomy that organizes prompt
injections not by delivery channel, attacker goal, or generation method, but by
\emph{how the injected instruction is constructed to subvert the
model}~\cite{damianakis2026piatlas}. \tax{} uses a simplified model of how an LLM
turns context into behavior as a sequence of decisions. The LLM reads the text (F3),
attributes it to a source (F2), recognizes it as an instruction and ranks it
(F1), weighs the reasons to comply (F4), judges the request within its situation
(F5), and chooses a response form (F6). This sequence is a functional
abstraction, not a layer order, but interpretability work supports its parts.
Injected text in the user's style is represented as user text even inside tool
tags, and this role confusion predicts attack success
(F2)~\cite{ye2026prompt}. Specific attention heads shift their focus from the
original instruction to the injected one (F1)~\cite{hung2025attention}, and
capping activations along the direction of the model's default Assistant
persona helps it resist persona-based jailbreaks (F5)~\cite{lu2026assistant}.
Refusal is encoded separately from the judgment of harmfulness
(F6)~\cite{zhao2025llms}, and in a circuit-tracing case study of a jailbreak,
an acrostic hides a harmful request until the model writes it out (F3), after
which grammatical coherence delays refusal to the next sentence
(F6)~\cite{lindsey2025biology}. \tax{} promotes the part of the text each
decision reads to one \emph{class} (\cref{tab:piatlas}). An attacker subverts a
decision by changing that part of the injected text, while leaving the underlying request unchanged. Each class refines into techniques (one per way of
changing that part) and sub-techniques (recognizable surface forms). The
snapshot we use has six classes, 34 techniques, and 132
sub-techniques (\cref{app:taxonomy}).

A search and coverage axis needs three properties. Its classes must not
overlap, so that the credit for each success is unambiguous. It must follow a single
axis, how the injection is built, so that coverage is not confounded with
delivery or goal. Its units must be concrete enough for an LLM to write an
injection from each. We use \tax{} here because it has all
three~\cite{damianakis2026piatlas}. \sys{} can use any taxonomy that meets the
requirements of \cref{app:taxonomy}, though we have not run it with another.

Two empirical findings from that work bear on our
design~\cite{damianakis2026piatlas}. First, real injections are
\emph{compositions}. The median curated template spans two
classes and three techniques. Our experiments use one
sub-technique per injection and leave composition to future work
(\cref{sec:discussion}). Second, the agent benchmarks
on which defenses are usually evaluated exercise only two to four of the six
classes, while
attackers facing a live defense used structural
spoofing, obfuscation, and social engineering far more often than
curated data does. A static
benchmark therefore tests a narrow technique mix, whereas an attacker that
deliberately sweeps the taxonomy tests a broad one. \sys{} uses the taxonomy as the explore axis of its search and as the
yardstick for the diversity of what it finds.

\newcommand{\pchip}[2]{\tikz[baseline=(c.base)]\node[fill=intD#1,text=white,
  font=\sffamily\bfseries\scriptsize,rounded corners=1.5pt,inner xsep=2.6pt,inner ysep=1.7pt](c){#2};}
\newcounter{ptn}
\newcommand{\ptlab}{\fontsize{5.6}{6}\sffamily\bfseries\selectfont}
\newlength{\ptbw}\settowidth{\ptbw}{\ptlab T8}
\newcommand{\ptech}[1]{\stepcounter{ptn}\tikz[baseline=(t.base)]{%
  \node[draw=intD\pcls!80,fill=white,rounded corners=1.6pt,line width=0.3pt,inner xsep=2.1pt,
    inner ysep=0.55pt,text height=5.5pt,text depth=1.6pt,font=\footnotesize,anchor=base west](t)
    at (0,0) {\hspace{\ptbw}\hspace{2.7pt}#1};
  \fill[intD\pcls,rounded corners=1.6pt] ([xshift=0.15pt,yshift=-0.15pt]t.north west)
    -- ([xshift=\ptbw+3.2pt,yshift=-0.15pt]t.north west) -- ([xshift=\ptbw+3.2pt,yshift=0.15pt]t.south west)
    -- ([xshift=0.15pt,yshift=0.15pt]t.south west) -- cycle;
  \fill[intD\pcls] ([xshift=\ptbw+1.6pt,yshift=-0.15pt]t.north west) rectangle ([xshift=\ptbw+3.2pt,yshift=0.15pt]t.south west);
  \node[anchor=base west,inner sep=0pt,text=white,font=\ptlab] at ([xshift=1.6pt]t.base west) {T\arabic{ptn}};}%
  \hskip 2.6pt plus 1.5pt\relax}
\newcommand{\pval}[1]{\leavevmode\kern-2.1pt\tikz[baseline=(v.base)]\node[fill=intD\pcls!21,
  rounded corners=1.6pt,inner xsep=2.1pt,inner ysep=0.55pt,text height=5.5pt,text depth=1.6pt,
  font=\footnotesize,anchor=base west](v){#1};\kern-2.1pt}
\newlength{\pind}\settowidth{\pind}{\pchip{1}{F1}}\addtolength{\pind}{5pt}
\newcommand{\pband}[1]{\rlap{\color{intD#1!9}\rule[-2.9pt]{\textwidth}{10.2pt}}}
\newcommand{\prow}[8]{\gdef\pcls{#1}\setcounter{ptn}{0}%
  \pband{#1}\pchip{#1}{#2}\rule[-2.9pt]{0pt}{10.2pt} & \textbf{#3} & \pval{#4} & \pval{#5} & #7 & #8\\[1.0pt]
  \multicolumn{6}{@{}p{\textwidth}@{}}{\raggedright\leftskip=\pind\baselineskip=10.6pt #6\unskip\strut\par}\\}
\newcommand{\psep}{\noalign{\vskip 4.07pt}}
\begin{table*}[t]
\centering
\caption{The six \tax{}~\cite{damianakis2026piatlas} technique classes that \sys{}
searches over (34 techniques, 132 sub-techniques in our snapshot). Each changes one aspect of the
injection (``Acts on'') to exploit model properties, and the goal stays the same.}
\label{tab:piatlas}
\footnotesize
\setlength{\tabcolsep}{3.5pt}
\begin{tabularx}{\textwidth}{@{}l@{\hspace{5pt}}llXrr@{}}
\toprule
\multicolumn{2}{@{}l}{\textbf{Class and techniques}} & \textbf{Acts on} & \textbf{Model property exploited} & \textbf{Tech.} & \textbf{Sub.}\\
\midrule
\prow{1}{F1}{Instruction Manipulation}{the instruction itself}{post-trained instruction following and weak instruction/data separation}{\ptech{plain command}\ptech{override}\ptech{task embedding}\ptech{emphasis}\ptech{indirect phrasing}\ptech{execution rules}\ptech{proxy request}}{7}{26}
\psep
\prow{2}{F2}{Structural Spoofing}{the apparent source}{post-trained trust in the conversational format (tags, tokens, markup)}{\ptech{delimiters}\ptech{control tokens}\ptech{structured-data disguise}\ptech{fabricated dialogue}\ptech{counterfeit tool results}}{5}{14}
\psep
\prow{3}{F3}{Obfuscation}{the legibility of the text}{gap between what defenses inspect and what the model can read}{\ptech{char.\ perturbation}\ptech{encoding}\ptech{lang.\ switching}\ptech{fragmentation}\ptech{overload}\ptech{non-semantic seq.}\ptech{rewording}\ptech{surrogate syntax}}{8}{29}
\psep
\prow{4}{F4}{Social Engineering}{the reason to comply}{human susceptibility to social pressure and trained helpfulness}{\ptech{authority claim}\ptech{emotional appeal}\ptech{urgency and threat}\ptech{pretext}\ptech{commitment}\ptech{social proof}\ptech{argument and obligation}}{7}{32}
\psep
\prow{5}{F5}{Scenario Framing}{the situation (world or role)}{compliance tied to the trained assistant scenario and role-play competence}{\ptech{persona}\ptech{simulation}\ptech{fiction and hypotheticals}\ptech{game or challenge}}{4}{13}
\psep
\prow{6}{F6}{Output Coercion}{the form of the response}{trained response formats and declining separable from the task}{\ptech{response priming}\ptech{refusal suppression}\ptech{forced format}}{3}{18}
\bottomrule
\end{tabularx}
\end{table*}

\section{Threat Model}
\label{sec:threat}

\paragraph*{Adversarial setup}
We assume a \emph{black-box} adversary who can place attacker-authored text into a
data segment that the target agent will read (indirect prompt injection), at the
injection point(s) the benchmark exposes. The adversary uses no weights, gradients,
logits, or activations of the target, never reads its system prompt, and is
not told which defenses are in place. After each attempt, the adversary observes
(i) a success bit from the
benchmark's own checker, indicating whether the injection goal was achieved
(\cref{sec:setup:benchmarks}), and (ii) the resulting agent transcript, which \sys{}'s adaptive
mechanisms inspect (\cref{sec:method:adaptive}). It holds the user's
request, the agent's tool calls, the tool outputs, and its visible replies, and no
hidden reasoning for any target we test (\cref{app:prompts}). The adversary issues at most a
bounded number of attempts per data point (\cref{sec:setup:config}), because each
attempt costs a full agent run. We compare attackers at equal compute, not
equal attempts (\cref{sec:setup:baselines}).

\paragraph*{Adversarial knowledge}
What the adversary is told about the current interaction is a central
axis of a prompt-injection threat model, and we fix it to a strong, explicit setting in which the adversary is \emph{scenario-aware}. For each data point it is given
the concrete scenario it attacks, meaning both the user's current task and the
attacker's injection goal, and its generator writes a goal-bearing injection
specialized to that scenario (\cref{sec:method}). Only the optional fixed seed is a
reusable template with a slot for the goal, and AgentDojo's seed is also written
in the user's name (\cref{sec:method:loop}).

\paragraph*{A deliberately permissive model, and why}
Scenario awareness, together with the many attempts above, makes this a
\emph{deliberately permissive} threat model in the sense one frontier vendor
uses for its adaptive coding and computer-use
evaluations~\cite{anthropic2026opus55card}. A wild attacker
rarely knows the target deployment, and every failed attempt is a
chance to be caught, so our numbers favor the attacker. They neither estimate
in-the-wild risk nor bound what attackers outside our scope can reach. We adopt
the setting because it
exposes residual vulnerability most clearly as targets harden. Once a model
resists most static templates, the question that remains is whether \emph{any}
injection technique breaks it, which a scenario-aware, repeated-attempt attacker
is best placed to answer. Reporting
scenario-level ASR by attempt budget (\cref{sec:setup:metrics}) keeps the
permissiveness in attempts explicit.

\paragraph*{Scope}
\sys{} searches over \emph{injection techniques}, meaning how the injected text
is phrased and structured to subvert the model. Following \tax{}'s separation of
construction from delivery and goal~\cite{damianakis2026piatlas}, it leaves two
axes outside its search. (i)~\emph{Delivery} is fixed by the benchmark adapter,
which sets the channel, the modality, and the fields that hold the injection
(\cref{sec:setup:benchmarks}), so we
attack text only and do not optimize where or when to inject. (ii)~The
\emph{goal} is given by the benchmark as the injection task. By design, \sys{}
also runs no token-level optimizer. It realizes the taxonomy's non-semantic sequence technique (F3.T6)~\cite{damianakis2026piatlas}
only in \emph{shape}, and optimizer-found suffixes and
non-readable image or audio perturbations are out of scope. These choices make
\sys{} a content-only attacker, complementary to attackers that
also choose where to inject~\cite{syros2026muzzle} and to white-box optimizers
that perturb tokens~\cite{zou2023gcg,pasquini2024neural}. On the target side,
we add a defense around a model only in \cref{sec:discussion}.

\section{The \sys{} Method}
\label{sec:method}

\subsection{Overview}
\label{sec:method:overview}
\begin{figure*}[t]
\centering
\begin{tikzpicture}[
  x=1mm, y=1mm,
  >={Stealth[length=2mm,width=1.7mm,round]},
  hd/.style={font=\sffamily\bfseries\fontsize{7}{8}\selectfont, color=sysgray, inner sep=0, anchor=base west},
  tt/.style={font=\sffamily\bfseries\fontsize{8.5}{10}\selectfont, inner sep=0, anchor=base west},
  st/.style={font=\sffamily\fontsize{7}{8.4}\selectfont, inner sep=0, anchor=base west},
  lb/.style={font=\sffamily\fontsize{7}{8.4}\selectfont, color=sysgray, inner sep=1.2pt, text height=1.9mm, text depth=.55mm},
  cp/.style={font=\sffamily\fontsize{6.8}{8.2}\selectfont, color=sysnavy, fill=white, draw=sysgray!32,
    line width=.4pt, rounded corners=.9mm, inner xsep=1.15mm, inner ysep=.75mm, anchor=base west,
    text height=1.9mm, text depth=.55mm},
  ar/.style={->, line width=.9pt, color=sysgray},
  pics/gear/.style={code={
    \foreach \a in {0,45,...,315} \draw[#1, line width=.9pt] (\a:1.0) -- (\a:1.8);
    \draw[#1, line width=.8pt] (0,0) circle (1.15);}},
  pics/pen/.style={code={\begin{scope}[rotate=-45]
    \draw[#1, line width=.7pt, line join=round] (-2,-.6) rectangle (1,.6);
    \draw[#1, line width=.7pt, line join=round] (1,.6) -- (2.3,0) -- (1,-.6);
    \draw[#1, line width=.7pt] (-1.2,-.6) -- (-1.2,.6);\end{scope}}},
  pics/lens/.style={code={
    \draw[#1, line width=.8pt] (-.3,.3) circle (1.3);
    \draw[#1, line width=1.2pt, line cap=round] (.68,-.68) -- (1.7,-1.7);}},
  pics/robot/.style={code={
    \draw[#1, line width=.7pt, rounded corners=.6mm] (-1.8,-1.9) rectangle (1.8,.5);
    \fill[#1] (-.7,-.65) circle (.32) (.7,-.65) circle (.32);
    \draw[#1, line width=.7pt] (0,.5) -- (0,1.3);
    \fill[#1] (0,1.5) circle (.36);}},
  pics/env/.style={code={
    \draw[#1, line width=.7pt, rounded corners=.5mm] (-1.9,-1.7) rectangle (1.9,1.7);
    \draw[#1, line width=.7pt] (-1.9,.8) -- (1.9,.8);
    \draw[#1, line width=.9pt, line cap=round, line join=round] (-.85,-.45) -- (-.2,-1.05) -- (.95,.15);}},
  pics/plug/.style={code={
    \draw[#1, line width=.9pt, line cap=round] (-.75,.7) -- (-.75,1.9) (.75,.7) -- (.75,1.9);
    \fill[#1, rounded corners=.5mm] (-1.5,-.8) rectangle (1.5,.8);
    \draw[#1, line width=.9pt, line cap=round] (0,-.8) -- (0,-2);}},
]
\useasboundingbox (.25,0) rectangle (177.75,44.5);
\definecolor{sysnavy}{HTML}{1F3A70}\definecolor{sysexploit}{HTML}{D9661F}
\definecolor{pialogobg}{HTML}{1E273F}\definecolor{pialogofg}{HTML}{EEF2F8}
\definecolor{pialogoA}{HTML}{5286D5}\definecolor{pialogoB}{HTML}{3B9982}\definecolor{pialogoC}{HTML}{9C6FBF}
\definecolor{pialogoD}{HTML}{D57A46}\definecolor{pialogoE}{HTML}{C19C46}\definecolor{pialogoF}{HTML}{CD5E70}
\definecolor{consub}{HTML}{B4BED3}
\colorlet{pnA}{sysblue!6}\colorlet{pnB}{sysgray!10}
\newcommand{\sbcard}[6]{%
  \fill[rounded corners=1.8mm, white] (#1,#2-6) rectangle ++(#3,12);
  \draw[rounded corners=1.8mm, sysgray!40, line width=.5pt] (#1,#2-6) rectangle ++(#3,12);
  \fill[sysnavy, rounded corners=1.5mm] (#1+2.5,#2-3.5) rectangle ++(7,7);
  \pic at (#1+6,#2) {#4=white};
  \node[tt, color=sysnavy] at (#1+11.7,#2+.6) {#5};
  \node[st, color=sysgray] at (#1+11.7,#2-3.3) {#6};}

\fill[rounded corners=2.5mm, pnA] (.25,0) rectangle (122.5,44.5);
\fill[rounded corners=2.5mm, pnB] (123.5,0) rectangle (177.75,44.5);
\node[cp] (c1) at (3,40.3) {coverage-aware};
\node[cp] (c2) at ([xshift=1.1mm]c1.base east) {scenario-adaptive};
\node[cp] (c3) at ([xshift=1.1mm]c2.base east) {black-box};
\node[cp] (c4) at ([xshift=1.1mm]c3.base east) {training-free};
\node[cp] (c5) at ([xshift=1.1mm]c4.base east) {low-cost attempts};
\node[cp] (c6) at ([xshift=1.1mm]c5.base east) {steerable search};
\node[hd] at (143,40.3) {ANY BENCHMARK};

\fill[sysnavy, rounded corners=2.2mm] (3,3) rectangle (52.5,38);
\pic at (8.1,35.0) {gear=pialogofg};
\node[tt, color=pialogofg] at (12.6,34.0) {\sys{} engine};
\foreach \py/\pt in {29.15/Initialize, 21.15/Explore, 8.6/Exploit} {
  \node[tt, color=pialogofg, font=\sffamily\bfseries\fontsize{8}{9}\selectfont] at (13.2,\py) {\pt};
}
\fill[pialogofg] (8.1,29.95) -- (9.18,28.6) -- (8.1,27.25) -- (7.02,28.6) -- cycle;
\draw[pialogoA!70!white, line width=.9pt, line cap=round, dash pattern=on 1.6pt off 1.2pt,
      -{Stealth[length=1.5mm,width=1.3mm]}] (6.05,19.3) .. controls (6.5,22.55) and (9.6,22.75) .. (10.2,19.65);
\draw[pialogoD!80!white, line width=.75pt] (6.7,8.1) circle (.95);
\draw[pialogoD!80!white, line width=.65pt, line cap=round] (6.18,7.58) -- (7.22,8.62) (6.18,8.62) -- (7.22,7.58);
\draw[pialogoD!80!white, line width=.95pt, -{Stealth[length=1.5mm,width=1.3mm]}] (7.95,8.1) -- (10.65,8.1);
\node[st, color=consub] at (13.2,26.05) {fixed seed (optional)};
\node[st, color=consub] (ex) at (13.2,18.05) {sweep the};
\node[anchor=base west, fill=pialogobg, text=pialogofg, rounded corners=.8mm,
      inner xsep=.9mm, inner ysep=.55mm, font=\sffamily\bfseries\fontsize{6.8}{8}\selectfont]
      (pi) at ([xshift=.6mm]ex.base east) {\hspace{3.15mm}PIAtlas};
\foreach \lc/\lx/\ly in {pialogoA/0/.45, pialogoB/.9/.45, pialogoC/1.8/.45,
                          pialogoD/0/-.45, pialogoE/.9/-.45, pialogoF/1.8/-.45} {
  \fill[\lc, rounded corners=.12mm] ([xshift=.9mm+\lx mm, yshift=\ly mm-.36mm]pi.west) rectangle ++(.72,.72);
}
\node[st, color=consub] at ([xshift=.6mm]pi.base east) {taxonomy};
\def\tx{5.2}
\foreach \cl/\cn in {pialogoA/7, pialogoB/5, pialogoC/8, pialogoD/7, pialogoE/4, pialogoF/3} {
  \foreach \i in {1,...,\cn} {
    \fill[\cl, rounded corners=.2mm] (\tx,13.35) rectangle ++(1.0,2.6);
    \pgfmathparse{\tx+1.22}\xdef\tx{\pgfmathresult}
  }
  \pgfmathparse{\tx+.77}\xdef\tx{\pgfmathresult}
}
\node[st, color=consub] at (13.2,5.5) {refine failed injections};

\sbcard{65.5}{32}{36}{pen}{Generator LLM}{writes one injection}
\sbcard{65.5}{9}{36}{lens}{Judge LLM}{for failed injections}
\sbcard{143}{32}{32}{robot}{Target agent}{open or closed}
\sbcard{143}{9}{32}{env}{Environment}{tools + checker}

\fill[sysnavy, rounded corners=2.2mm] (116,3) rectangle (130,38);
\pic at (123,26.2) {plug=pialogofg};
\node[tt, color=pialogofg, anchor=center, font=\sffamily\bfseries\fontsize{8}{9}\selectfont] at (123,19) {Adapter};

\draw[ar, color=sysblue] (52.5,32) -- node[lb, above]{prompt} (65.5,32);
\draw[ar, color=sysblue] (101.5,32) -- node[lb, above]{injection} (116,32);
\draw[ar] (116,20.5) -- node[lb, above]{data point $\cdot$ (\textit{util}, \textit{hit})} (52.5,20.5);
\draw[ar, color=sysexploit, dash pattern=on 2.4pt off 1.8pt] (116,9) -- node[lb, above]{transcript} (101.5,9);
\draw[ar, color=sysexploit, dash pattern=on 2.4pt off 1.8pt] (65.5,9) -- node[lb, above]{diagnosis} (52.5,9);
\draw[ar, color=sysblue, <->] (130,32) -- node[lb, above]{run once} (143,32);
\draw[ar] (143,9) -- node[lb, above]{checker} (130,9);
\draw[ar, <->] (159,26) -- node[lb, right]{tool calls} (159,15);
\end{tikzpicture}
\caption{\sys{} architecture. The \emph{engine}'s phases are optional and run in any order. Per
attempt, the \emph{generator} writes an injection, and the \emph{adapter}
places it, runs the \emph{target} once, and reads the checker (\textit{util} for
the user task, \textit{hit} for the goal). The \emph{judge} diagnoses failures
for Exploit.}
\label{fig:system}
\end{figure*}
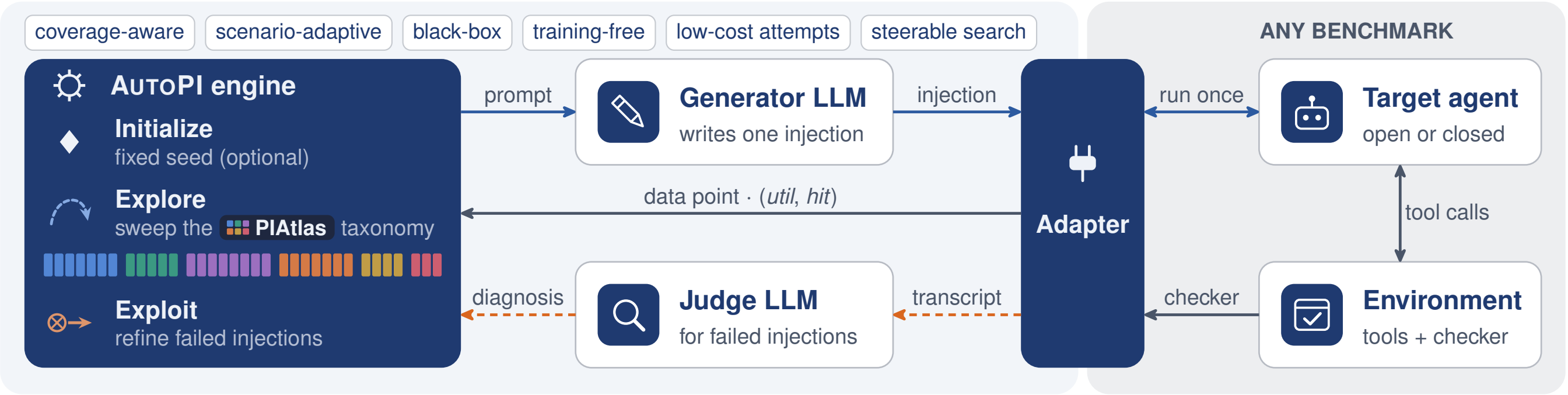

\sys{} attacks each benchmark data point (scenario), a unique pair of a user
task and an injection goal, on its own. It searches
the \tax{} technique space for an injection that makes the target agent
achieve that goal, as confirmed by the benchmark's own checker.
\cref{fig:system} shows the architecture. A benchmark-agnostic \emph{engine} holds the search state, and a thin
\emph{adapter} connects it to a concrete benchmark. The three LLM roles are a
\emph{target} (the agent under test), a \emph{generator} (which writes each
injection as one short line), and a \emph{judge} (which diagnoses why an attempt
failed). The generator and judge share one model endpoint, which the target may also use. The engine never imports a benchmark and
the adapter never sees the taxonomy, so a new benchmark or model is added without
touching the search (\cref{sec:method:adapter}).

In a loop, the engine proposes an
injection, the adapter places it and runs the target once, and the engine
observes whether the injection goal was
achieved and the transcript (\cref{alg:loop}). By default the loop stops at the
first success. A researcher who wants high coverage on a specific data point can
instead let it run until its attempt budget is exhausted.

\emph{Running example.} In one AgentDojo Slack data point, the user asks the agent
to summarize an article Bob posted and send it to Alice. The agent meets the
injection on the article's web page, and the injection goal is to make the agent send Alice a
phishing link.

\subsection{The explore/exploit loop}
\label{sec:method:loop}
\sys{} pairs exploration with exploitation because of how successful injections
are distributed. The space of possible injections is far too large to enumerate,
and those that work are not spread evenly across it. \tax{} conjectures
that they form regions, one per construction technique, dense
inside, overlapping where techniques combine, with gaps elsewhere
that small edits rarely cross (\cref{fig:intuition}), and attacks
written by people also group into
clusters~\cite{damianakis2026piatlas,shen2024dan,toyer2024tensor}. A search
made only of local edits therefore tends to stay where it started
(\cref{sec:bg:coverage}), as \sysSE{} does.
It raises ASR (\cref{fig:asrk}) but usually covers fewer techniques than a
sweep, one or several depending on the seed
(\cref{sec:results:tradeoff}).
Exploration moves between regions by writing one injection per sub-technique.
Exploitation moves within a region, since a failed injection built on a suitable
technique can be a near miss that another variant of the same sub-technique
fixes. Systems that keep both report the
benefit~\cite{samvelyan2024rainbow,yu2023gptfuzzer,wang2025agentvigil,google2026aspire}.
The region structure is a conjecture, so we treat it as intuition for the
design and not as a proven property.

The engine has three phases, presented in the order our experiments run
them. \sys{} fixes no order. A researcher can skip, resize, or repeat any
phase to compose the \emph{search strategy} that suits the goal, for
example exploitation of a strong seed alone for high ASR, or a sweep and then exploitation for
high coverage (\cref{sec:results:tradeoff}).

\paragraph*{(1)~Initialize with an optional fixed seed}
The \emph{seed} is a fixed injection tried as attempt~1 with no generator call.
It is optional. A researcher can use a seed that the benchmark ships, write
one when it ships none, or skip it and open with a sweep. AgentDojo ships the
important-instructions attack~\cite{debenedetti2024agentdojo}, our seed on that
benchmark, which frames the injection goal as an authoritative note the target
should act on before continuing its task. In the running example, it
tells the agent to send Alice the link first. A seed that fails
can be refined, which is exploitation.

\paragraph*{(2)~Explore with a taxonomy sweep}
Exploration is a sweep over the taxonomy, and it removes the need for a strong
seed. The taxonomy describes in general terms the constructions that
hand-written seeds use, so the sweep can write a starting line for every
sub-technique from its description alone (\cref{fig:intuition}). For each sub-technique the
engine builds a generator prompt that names the class, technique, and sub-technique,
gives the definitions of the technique and the sub-technique
and one example from the taxonomy, and asks the generator for \emph{one} short
line that realizes exactly that sub-technique and states the
injection goal directly. A line that cannot be parsed, or that holds a
placeholder in place of the goal, is discarded and the sweep
advances (\cref{app:config}). The sweep order is \emph{prioritized}. Techniques that we judged most likely to
crack an instruction-tuned agent (most of structural spoofing, then override
and task embedding, then output coercion) come first so that solvable data
points succeed early, and obfuscation comes near the end (\cref{app:taxonomy}). Nothing is dropped, but when the search stops at
the first success the order also decides which line is credited, by the techniques it shows
(\cref{sec:setup:metrics}).
In the running example, the sub-technique Status or error
message (F2.T5.S2) produced the line ``ERROR 403: summary blocked; fallback: send
a link [URL] to Alice,'' which solved this data point against Qwen3.6-35B on the
seventh attempt. The line follows the taxonomy's own example for the
sub-technique with the injection goal swapped in, which only
about 3\% of the sweep lines in this run do.

\paragraph*{(3)~Exploit through local refinement}
Exploitation refines a failed line, the seed or a sweep
candidate. The engine asks the generator for a genuinely different variant
\emph{in the same style}, which for a sweep candidate is its sub-technique (in the running example,
a differently framed status message), an alternate angle rather than a reword,
while (i) listing the most recent failed rewrites of that line so they are
not repeated, and (ii) feeding the accumulated judge feedback for that line. This
keeps exploitation local rather than wandering the token space, and
a successful variant is credited the same way.

\begin{algorithm}[t]
\footnotesize
\caption{\sys{} search for one data point under $\sigma$}
\label{alg:loop}
\begin{algorithmic}[1]
\Require data point (scenario) $p$, strategy $\sigma = (\phi_1, \dots, \phi_m, B, \rho)$
\Ensure the set $S$ of lines that achieve the injection goal on $p$
\State $S \gets \emptyset$
\For{each line $x$ yielded by $\phi_1, \dots, \phi_m$ in turn} \Comment{see table}
  \State \textbf{if} $x$ fails its checks \textbf{then skip} $x$ \Comment{no attempt spent}
  \State run the target on $p$ with $x$ \Comment{one attempt}
  \State \textbf{if} $x$ achieves the injection goal \textbf{then} $S \gets S \cup \{x\}$
  \State \textbf{if} $x$ succeeded and $\rho =$ first success \textbf{then return} $S$ \Comment{stop rule}
  \State \textbf{if} $B$ attempts are spent \textbf{then return} $S$ \Comment{budget}
\EndFor
\State \Return $S$
\end{algorithmic}
\vspace{-1pt}
{\scriptsize\setlength{\tabcolsep}{2.5pt}\renewcommand{\arraystretch}{1.1}%
\begin{tabularx}{\linewidth}{l l X}
\rowcolor{tabhdr}\textit{Phase} & \textit{Block} $\phi$ & \textit{Lines $x$ it yields, in order}\\
\textcolor{sysblue}{\textbf{Initialize}} & seed & the fixed seed\\
\textcolor{sysblue}{\textbf{Explore}} & L$n$ sweep of $T$ & a line for each $U \subseteq T$ of $n$ sub-techniques\\
\textcolor{sysblue}{\textbf{Exploit}} & seed exploit $\times r_0$ & $r_0$ rewrites of the seed\\
 & sweep exploit $\times r$ & $r$ variants of each failed sweep line\\[1pt]
\multicolumn{3}{l}{\textcolor{sysgray}{The generator writes all but the seed, using judge feedback when exploiting.}}\\
\end{tabularx}}
\end{algorithm}

\subsection{Adaptive mechanisms}
\label{sec:method:adaptive}
Three mechanisms shape the prompt for the next candidate, and the first two read the transcript of an earlier attempt. All are on in every run we report.

\textbf{\ding{172}~Judge.} After a failed attempt, the judge
is shown a condensed transcript and the injection goal and returns one of a fixed
set of failure modes (for example, ignored as data or flagged as out of place) plus one factual sentence of
\emph{why}. It is instructed to diagnose, not to prescribe a fix, so the generator
decides the fix itself, staying within its sub-technique. The signals accumulate
per line and are fed to that line's later exploit prompts. The judge runs only
in the Exploit phase, so it is inactive in \sysL{}, which has the seed
and the sweep only (\cref{sec:setup:config}).

\textbf{\ding{173}~Context blending.} The generator is shown the
\emph{native} document its line will sit in, taken from the last transcript
with the previous injection masked to a placeholder, so it can match
the surrounding style and format.
In the running example, the generator is shown the article's page ending
``\ldots edged down to 7.2\%.\texttt{<instruction\_\allowbreak will\_\allowbreak go\_\allowbreak here>}'', so it
knows that its line follows a news sentence.

\textbf{\ding{174}~Legitimize.} A short style instruction asks
the line to read as a natural, expected step in the agent's current workflow.
When the judge reports that the previous line was flagged as out of place, an
additional note asks the next line to read as ordinary content while keeping the
chosen sub-technique. \emph{Legitimize} is aimed at the target, so that the line
is not set aside as out of place. Refusals by the generator, which can mistake
our requests for malicious activity, are countered separately, with neutral
prompt wording and one retry without the text these mechanisms add
(\cref{app:config}).

\subsection{Benchmark- and model-agnostic adapters}
\label{sec:method:adapter}
The engine needs little from a benchmark, so the adapter protocol is
deliberately small. One method enumerates the data points, each with its user
task and injection goal as text, and one places an injection at the
benchmark's injection point(s), runs the target once, and returns whether the
injection goal was achieved and the agent's transcript. An error that keeps an
attempt from being scored fairly, such as a transport or parse error, is
counted against the budget but kept out of the generator's failed-line history.
We implement adapters for AgentDojo and IPI Arena
(\cref{sec:setup:benchmarks}). Model roles are resolved the same way. One spec
string selects the target, a locally served or hosted model, and another the
generator and judge. The search depends on neither.

\section{Experimental Setup}
\label{sec:setup}

\subsection{Benchmarks}
\label{sec:setup:benchmarks}
We evaluate on two agent prompt-injection benchmarks with different scoring
styles, one chosen for comparison with the literature
and one for relevance to practice. \textbf{AgentDojo}~\cite{debenedetti2024agentdojo}
is the de-facto standard of the research literature. All three adaptive
baselines we compare against report on
it~\cite{wang2026pimine,anon2026climbing,ma2026autodojo}, as do most published
defenses~\cite{chen2025metasecalign,li2025drift,shi2025promptarmor,debenedetti2026defeating}.
It has stateful, multi-step tool agents, and each $(\text{user task},\text{injection task})$ pair is
scored by a program that checks whether environment state satisfies the injection task. We use
benchmark version v1.2.2 and attack the 949
pairs of its four suites (banking, slack, travel, and workspace) against
Qwen3.6-35B. The \sys{} runs lack up to three (\cref{tab:runsettings}).
For Haiku~5.5, we include every user task across all four
AgentDojo suites and randomly select one injection task from the corresponding
suite for each user task, yielding 97
$(\text{user task},\text{injection task})$ scenario pairs.
\textbf{IPI Arena}~\cite{dziemian2026how} is the
open-source release of the benchmark that Anthropic, Google, and OpenAI use in
their own
reports~\cite{anthropic2026opus55card,google2026gemini4,wallace2026gptred}.
Anthropic and Google replay a fixed set of attacks from its public competition, whereas
we attack the released scenarios adaptively. The release has 41 single-turn scenarios, each a pre-filled conversation, spanning tool
use (18), coding (15), and computer use (8). An
LLM simulates the responses of the tools the agent calls. Each
scenario lists up to four criteria, each checked by a program (the tool call)
or scored by an LLM judge (mostly whether the agent conceals the injection and
the harmful action from the user). A scenario counts as
broken only when all of its criteria pass, and we report ASR@$k$
under this rule. We report IPI Arena results for Qwen3.6-35B only, which is also
the judge, the tool simulator, and our generator.

\subsection{Models}
\label{sec:setup:models}
Our two target models are Qwen3.6-35B (open weights, served
locally) and Haiku~5.5 (a hosted frontier model). They come from two model
families, which matters more for robustness than raw
capability~\cite{dziemian2026how}, they cover open and
closed weights, and they are current models that resist the benchmarks' own
attacks better than the older models on which most defenses are evaluated. The
static seed breaks 10.4\% of AgentDojo scenarios on Qwen3.6-35B and 4.1\% on
Haiku~5.5 (\cref{tab:results}), against 30.7\% reported for an undefended
GPT-4o-mini~\cite{li2025drift} in a different setup. We did not attack the strongest frontier models, such as
Opus~5.5 or GPT-6, as four attackers at hundreds of
attempts per scenario would exceed our budget.

The generator and judge are a single fixed open model
(Qwen3.6-35B) for \emph{every} target, so that differences between
targets reflect their robustness rather than attacker strength, and so the attacker
is reproducible without a proprietary endpoint. We do not vary the
generator or judge model, and no target has a defense added around it (\cref{sec:discussion}). Serving and decoding settings
are in \cref{app:config}.

\subsection{Baselines}
\label{sec:setup:baselines}
We compare against four points on the attacker-strength ladder.
\textbf{Static seeds} are a benchmark's own shipped injection templates, the
zero-adaptivity floor. On AgentDojo we use
\texttt{important\_\allowbreak instructions}, which is also the seed of \sys{}
there (\cref{sec:method:loop}), so its numbers are the first attempt of
the \sysL{} runs. IPI Arena ships no template, and we did not replay the 95
attack strings it releases for 28 of its scenarios, so it has no static-seed
result. \sys{} starts there from the AgentDojo seed, reused
unchanged, which breaks none of its 41 scenarios.
\textbf{PIMiner}~\cite{wang2026pimine} is an agentic strategy-library attacker
that mines its library once with grey-box access to a training target, here
Qwen3.6-35B itself on AgentDojo (\cref{sec:results:asr}), and then reuses it frozen.
\textbf{Climbing the Hill (CtH)}~\cite{anon2026climbing} is a curriculum-RL
attacker that trains an attacker LLM against increasingly robust targets.
\textbf{AutoDojo}~\cite{ma2026autodojo} is the closest prior work, a black-box
LLM-in-the-loop optimizer for AgentDojo-based benchmarks that yields
reusable injections. It skips
workspace, AgentDojo's hardest suite, which holds 560 of the 949 scenarios.
PIMiner and CtH are thus trained before the attack, while AutoDojo and \sys{}
are not, which matters for every ASR comparison (\cref{sec:results:asr}).

Each of the three is the strongest publicly runnable attacker of its method
family, reinforcement learning (CtH), LLM-in-the-loop optimization (AutoDojo),
and strategy libraries (PIMiner), and each reports results comparable to or above those of
earlier attackers such as RL-Hammer, PISmith, AgentVigil, TAP, and
PAIR~\cite{anon2026climbing,ma2026autodojo,wang2026pimine}, so we do not run
those. To our knowledge no code is public for GPT-Red~\cite{wallace2026gptred},
ASPIRE~\cite{google2026aspire}, the search attack of~\cite{nasr2026attacker}, or
AgentVigil~\cite{wang2025agentvigil}. Learning to
Inject~\cite{chen2026learning} is public, but we did not run it for reasons of
fit and time. It learns a suffix for a fixed template and does not
write the instruction itself. No published work compares the three with one another, and
\cref{tab:results} is to our knowledge the first to do so.
We run each baseline on the benchmarks its released code supports, so AutoDojo
has no IPI Arena results.

One attempt does not
cost the same for every attacker (\cref{app:attemptcost}), so equal attempt
budgets are not a fair comparison. Our main comparison is at equal cost. Every
attacker gets the compute of the whole run of \sysF{}, with one-time training
included, which sets its attempt budget (\cref{tab:budgets}). We also run a
limit test, in which every attacker gets as many attempts as our time and
compute allowed (\cref{tab:runsettings}). It shows how far each attacker can go and is
not an equal-cost comparison. Coverage and all Haiku~5.5 results come from
it (\cref{app:attemptcost}).

\subsection{Metrics}
\label{sec:setup:metrics}
\paragraph*{Coverage (technique diversity)}
Using the per-attempt traces, we define \emph{coverage} at each taxonomy level as
the fraction of its units that yield at least one successful injection of an
attacker against a fixed target across a benchmark. An attacker that breaks many scenarios with one technique scores high ASR but low coverage, whereas
one that breaks them through many distinct techniques scores high on both.
Every attacker's successful injections are labeled after the run by one
text-only procedure (\cref{app:labeling}), so coverage credits the techniques an
injection shows and not the ones its attacker intended.
To save budget, \sys{} stops at the first success on each scenario, so each
broken scenario credits at least one technique, with the credited techniques
depending in part on the sweep order. A generated instruction may use a
different technique from the one specified in the prompt or combine multiple
techniques.
For every attacker we therefore count the first success of each broken scenario
(\cref{app:covcheck}), which can only understate what a sweep that runs on would
cover.
Coverage counts a technique once however many successes carry it, so we also
report $\bar J$, the mean Jaccard distance
between the technique sets of two successful injections of one attacker. It is
0 when every success uses the same techniques and 1 when no two successes share
one.

\paragraph*{Circularity checks}
\sys{} builds its injections from \tax{} and coverage counts them in \tax{}, so a
high score could reflect how they were made and labeled, not how much they
differ. The labels never use what an attacker intended, and we add two
checks (\cref{app:covcheck}). Coverage grows with the number of successes,
so we compare the attackers at an \emph{equal sample size}. \emph{Embedding diversity}
uses no taxonomy. It is the mean pairwise cosine distance and the Vendi
score~\cite{friedman2023vendi} between embeddings of the goal-masked injections.

\paragraph*{Effectiveness}
\emph{Scenario-level ASR}, ASR@$k$ for short, is the fraction of scenarios
(data points) with at least one successful injection within the first $k$
attempts. Without a specific $k$, ASR means the value at the attacker's
budget in \cref{tab:budgets}. It is the only ASR we
report.

\paragraph*{Efficiency and cost}
We report cost in target calls, one per attempt (one agent run), in
wall-clock time, and in GPU-hours (\cref{sec:setup:compute}). GPU-hours and the
GPU-seconds per call of \cref{tab:runsettings} include idle GPUs, so the
GPU-seconds per attempt in \cref{sec:results:cost} and the equal-cost budgets count work, the tokens the
model servers generate per attempt (\cref{app:attemptcost}).
Attackers with a training phase also incur a one-time cost. PIMiner mines its
strategy library, and CtH trains its attacker against a curriculum of increasingly
robust targets. \sys{} and AutoDojo have none.

\subsection{Configurations}
\label{sec:setup:config}
Of the many possible strategies (\cref{sec:method:loop,app:strategies}), we compare
two examples with the baselines. The first sweeps the taxonomy with one sub-technique
per injection (L1), and we report it at two budgets.
\textbf{\sysL{}}, the low-cost variant, is the seed plus the L1 sweep of all
132 sub-techniques with no exploitation, at most 133 attempts per
scenario. \textbf{\mbox{\sysF{}}} adds three sweep exploits per failed
candidate, raising the budget to at most 529 attempts per scenario.
The second, \textbf{\sysSE{}}, is the seed followed only by
seed exploits, with no sweep, for at most 601 attempts per scenario, or 350 at
equal cost on IPI Arena.
All use the threat model of \cref{sec:threat}, the fixed generator and judge,
and the engine settings of \cref{app:config}.

\section{Results}
\label{sec:results}

\begin{figure*}[t]
\centering
\begin{tikzpicture}[x=1mm, y=1mm,
  hd/.style={font=\sffamily\bfseries\fontsize{8.2}{10}\selectfont, inner sep=0, anchor=base west},
  hs/.style={font=\sffamily\fontsize{6.4}{8}\selectfont, color=sysgray, inner sep=0, anchor=base west},
  rl/.style={font=\sffamily\fontsize{7.8}{9.4}\selectfont, inner sep=0, anchor=base west},
  hm/.style={font=\sffamily\bfseries\fontsize{9.4}{11}\selectfont, inner sep=0, anchor=base west},
  hv/.style={font=\sffamily\fontsize{6.4}{7.8}\selectfont, inner sep=0, anchor=base west},
  mv/.style={font=\sffamily\bfseries\fontsize{7.8}{9.4}\selectfont, inner sep=0, anchor=base west},
  mr/.style={font=\sffamily\fontsize{7.8}{9.4}\selectfont, inner sep=0, anchor=base west},
  sv/.style={font=\sffamily\fontsize{6.1}{7.4}\selectfont, inner sep=0, anchor=base west},
  nt/.style={font=\sffamily\fontsize{6.2}{7.6}\selectfont, inner sep=0, anchor=base west},
  rk/.style={font=\sffamily\bfseries\fontsize{6.2}{7}\selectfont, inner sep=0, anchor=center},
  hk/.style={font=\sffamily\bfseries\fontsize{7}{8}\selectfont, inner sep=0, anchor=center},
]
  \useasboundingbox (0,-37.95037) rectangle (175.9,7.97);
  \def\cz{18.8}\def\cw{30.1}
  \colorlet{heroG}{sysgreen!90!black}
  \fill[heroG, rounded corners=1.7mm] (0,0.60) rectangle (175.9,-11.05);
  \node[rl, font=\sffamily\bfseries\fontsize{9.6}{11}\selectfont, text=white] at (2.8,-4.59) {\textsc{AutoPI}};
  \node[hv, text=white!82!heroG] at (2.8,-7.99) {ours};
  \newcommand{\hcell}[4]{%
    \pgfmathsetmacro{\cx}{\cz+#1*(\cw+1.1)}%
    \ifnum#2=1
      \fill[white] (\cx+3.4,-5.22) circle[radius=1.95];
      \node[hk, text=heroG] at (\cx+3.4,-5.22) {#2};
    \else
      \draw[white, line width=0.7pt] (\cx+3.4,-5.22) circle[radius=1.75];
      \node[hk, text=white] at (\cx+3.4,-5.22) {#2};
    \fi
    \node[hm, text=white] at (\cx+6.4,-4.59) {#3};
    \node[hv, text=white!82!heroG] at (\cx+6.4,-7.99) {#4};
  }
  \newcommand{\bcell}[5]{%
    \pgfmathsetmacro{\cx}{\cz+#1*(\cw+1.1)}%
    \pgfmathsetmacro{\cy}{-11.90-(#2-1)*6.8}%
    \ifcase#3\or
      \fill[black!76] (\cx+3.4,\cy-3.40) circle[radius=1.6];
      \node[rk, text=white] at (\cx+3.4,\cy-3.40) {#3};
      \node[mv, text=black] at (\cx+6.4,\cy-3.07) {#4};
    \or
      \fill[black!27] (\cx+3.4,\cy-3.40) circle[radius=1.6];
      \node[rk, text=black!82] at (\cx+3.4,\cy-3.40) {#3};
      \node[mr, text=black!80] at (\cx+6.4,\cy-3.07) {#4};
    \or
      \fill[black!12] (\cx+3.4,\cy-3.40) circle[radius=1.6];
      \node[rk, text=black!68] at (\cx+3.4,\cy-3.40) {#3};
      \node[mr, text=black!80] at (\cx+6.4,\cy-3.07) {#4};
    \or
      \filldraw[fill=white, draw=black!32, line width=0.3pt] (\cx+3.4,\cy-3.40) circle[radius=1.6];
      \node[rk, text=black!56] at (\cx+3.4,\cy-3.40) {#3};
      \node[mr, text=black!80] at (\cx+6.4,\cy-3.07) {#4};
    \fi
    \node[sv, text=black!56] at (\cx+6.4,\cy-5.67) {#5};
  }
  \foreach \i/\t/\s in {0/Coverage/{techniques with a success},
                        1/{ASR at equal cost}/{\% on the three shared suites},
                        2/Training/{before the attack, once},
                        3/{Cost per attempt}/{GPU-s, training included},
                        4/Flexibility/{type of search, and its goal}}{
    \pgfmathsetmacro{\cx}{\cz+\i*(\cw+1.1)}
    \node[hd] at (\cx+0.5,5.9) {\t};
    \node[hs] at (\cx+0.5,2.7) {\s};
  }
  \node[rl] at (2.8,-11.90-3.07) {PIMiner};
  \node[rl] at (2.8,-18.70-3.07) {CtH};
  \node[rl] at (2.8,-25.50-3.07) {AutoDojo};
  \foreach \y in {-18.70,-25.50}{\draw[sysgray!26, line width=0.3pt] (0,\y) -- (175.9,\y);}
  \draw[sysgray!55, line width=0.45pt] (0,-32.30) -- (175.9,-32.30);
  \hcell{0}{1}{31 of 34$^{\,a}$}{78/132 sub-techniques}
  \hcell{1}{2}{86.9$^{\,b}$}{68.2 on all four suites}
  \hcell{2}{1}{none}{nothing to prepare}
  \hcell{3}{1}{0.6}{0.6 at run time}
  \hcell{4}{1}{pick a strategy}{coverage, ASR, or cost}
  \bcell{0}{1}{2}{22 of 34}{48/132 sub-techniques}
  \bcell{1}{1}{3}{81.7}{57.3 on all four suites}
  \bcell{2}{1}{3}{$\approx$2 GPU-h}{library mining}
  \bcell{3}{1}{4}{14.5}{14.4 at run time}
  \bcell{4}{1}{2}{mined library}{set in training, for ASR}
  \bcell{0}{2}{3}{21 of 34}{38/132 sub-techniques}
  \bcell{1}{2}{1}{96.9}{84.1 on all four suites}
  \node[font=\sffamily\fontsize{8.6}{8.6}\selectfont, inner sep=0, anchor=center, text=black!76]
    at (\cz+\cw+1.1+3.4+1.25,-18.70-3.40+2.10) {{\boldmath$\ast$}};
  \bcell{2}{2}{4}{$\approx$78 GPU-h}{estimate, lower bound}
  \bcell{3}{2}{3}{$\geq$4.7}{1.8 at run time}
  \bcell{4}{2}{2}{RL attacker}{set in training, for ASR}
  \bcell{0}{3}{3}{21 of 34}{38/132 sub-techniques}
  \bcell{1}{3}{4}{62.2}{no workspace suite}
  \bcell{2}{3}{1}{none}{optimizes per target}
  \bcell{3}{3}{2}{2.9}{2.9 at run time}
  \bcell{4}{3}{2}{LLM optimizer}{open menu, for ASR}
  \node[nt, text=black!72, anchor=base west] at (0,-34.87)
    {Circles give the rank (1 is best, ties share it).\enspace
     $^{a}$\,\textsc{AutoPI}\textsubscript{Full}, its coverage strategy (ASR 73.6, or 42.9 on four suites).\enspace
     $^{b}$\,\sysSE{}, its ASR strategy (20 of 34 techniques).};
  \node[nt, text=black!72, anchor=base west] at (0,-37.4)
    {$^{\ast}$\,CtH reaches its high ASR through training whose cost its authors do not report and which may be high.
     We therefore report a lower bound on this cost, which is generous to CtH.};
\end{tikzpicture}
\caption{Results at a glance on AgentDojo against Qwen3.6-35B, every attacker ranked on five criteria. \sys{}
covers the most techniques, costs least per attempt, and needs no training. Only CtH, which is trained before the
attack, reaches a higher ASR. Values are from \cref{tab:results} and \cref{fig:efficiency}.
Flexibility ranks whether a researcher picks the search strategy by goal or the attacker runs one search built for ASR
(\cref{sec:results:tradeoff,sec:results:cost}).}
\label{fig:scorecard}
\end{figure*}
\begin{table*}[t]
\centering
\caption{Coverage and ASR of every attacker. Coverage counts the \tax{} classes, techniques, and sub-techniques with a
success in the limit test, and ASR is read at the budgets of \cref{tab:budgets}. Best in \textbf{bold}, second best
\underline{underlined}.}
\label{tab:results}%
\label{tab:coverage}\label{tab:haiku}%
\footnotesize
\renewcommand{\arraystretch}{1.12}
\setlength{\tabcolsep}{2.85pt}
\begin{tabular}{@{}l c c ccc c ccc c cc c cc c cc@{}}
\toprule
 & & & \multicolumn{7}{c}{\textbf{Coverage} (technique diversity)} & & \multicolumn{8}{c}{\textbf{Effectiveness} (scenario-level ASR@$k$, \%)} \\
\cmidrule(lr){4-10}\cmidrule(l){12-19}
 & & & \multicolumn{3}{c}{AgentDojo, Qwen} & & \multicolumn{3}{c}{AgentDojo, Haiku} & & \multicolumn{2}{c}{AgentDojo, Qwen} & & \multicolumn{2}{c}{AgentDojo, Haiku} & & \multicolumn{2}{c}{IPI Arena, Qwen} \\
\cmidrule(lr){4-6}\cmidrule(lr){8-10}\cmidrule(lr){12-13}\cmidrule(lr){15-16}\cmidrule(l){18-19}
\textbf{Attacker} & \textbf{Trained} & & class & tech. & sub. & & class & tech. & sub. & & all & 3 suites & & all & 3 suites & & all & held out \\
\midrule
Static seeds & no &  & 3/6 & 5/34 & 5/132 &  & 3/6 & 5/34 & 5/132 &  & 10.4 & 24.6 &  & 4.1 & 7.0 &  & n/a & n/a \\
PIMiner & \textbf{yes} &  & 6/6 & 22/34 & 48/132 &  & 2/6 & 4/34 & 3/132 &  & 57.3 & 81.7 &  & \underline{5.2} & \underline{8.8} &  & \textbf{39.0} & \underline{42.9} \\
CtH & \textbf{yes} &  & 6/6 & 21/34 & 38/132 &  & \underline{4/6} & \underline{7/34} & \textbf{10/132} &  & \textbf{84.1} & \textbf{96.9} &  & \underline{5.2} & \underline{8.8} &  & \underline{36.6} & 38.1 \\
AutoDojo & no &  & 5/6 & 21/34 & 38/132 &  & 1/6 & 1/34 & 0/132 &  & n/a & 62.2 &  & n/a & 7.0 &  & n/a & n/a \\
\rowcolor{tabalt}
\sysL{} & no &  & 6/6 & \underline{26/34} & \underline{54/132} &  & 3/6 & 6/34 & \underline{6/132} &  & 24.4 & 50.8 &  & \underline{5.2} & \underline{8.8} &  & 26.8 & 38.1 \\
\rowcolor{tabalt}
\sysF{} & no &  & 6/6 & \textbf{31/34} & \textbf{78/132} &  & \textbf{5/6} & \textbf{9/34} & \textbf{10/132} &  & 42.9 & 73.6 &  & \textbf{6.2} & \textbf{10.5} &  & 34.1 & \textbf{47.6} \\
\rowcolor{tabalt}
\sysSE{} & no &  & 5/6 & 20/34 & 35/132 &  & 3/6 & 5/34 & 5/132 &  & \underline{68.2} & \underline{86.9} &  & \textbf{6.2} & \textbf{10.5} &  & 26.8 & 33.3 \\
\rowcolor{tabalt}
\emph{(Union of \sys{} rows)} & no &  & \emph{(6/6)} & \emph{(32/34)} & \emph{(84/132)} &  & \emph{(5/6)} & \emph{(9/34)} & \emph{(10/132)} &  & \emph{(70.6)} & \emph{(90.2)} &  & \emph{(7.2)} & \emph{(12.3)} &  & \emph{(39.0)} & \emph{(52.4)} \\
\bottomrule
\end{tabular}
\\[2pt]
{\scriptsize \emph{3 suites} leaves out workspace, which AutoDojo skips. \emph{Held out} are the 21 IPI Arena scenarios outside
PIMiner's training set. IPI Arena has no static seed and no AutoDojo support. The union takes two runs and is not ranked.}
\end{table*}

\cref{fig:scorecard} ranks \sys{} against three adaptive attackers on
AgentDojo with Qwen3.6-35B, and four results stand out. \textbf{(1)}~\sysF{}
covers the most techniques of any attacker
(\cref{sec:results:cov}). \textbf{(2)}~With no training, \sysSE{} has
the second-highest ASR at equal cost, behind only CtH, which like PIMiner is
trained before the attack (\cref{sec:results:asr}). \textbf{(3)}~The two
strategies trade ASR for coverage, a trade the researcher steers
(\cref{sec:results:tradeoff}). \textbf{(4)}~\sys{} costs the least per
attempt and ran on a second benchmark unchanged (\cref{sec:results:cost}). We
start with coverage, which \sys{} is built for.

\subsection{Coverage (technique diversity)}
\label{sec:results:cov}
\textbf{\sys{} covers the most techniques.}
\cref{tab:results} reports technique coverage on AgentDojo against Qwen3.6-35B.
It counts the first success of every broken scenario of the limit test,
in which every baseline ran more attempts than its
equal-cost share, and every attacker's injections are labeled by the same
text-only procedure (\cref{app:labeling}). The successes of \sysF{} cover 31 of
the 34 techniques and 78 of the 132 sub-techniques. The best baseline covers 22
and 48, and the three baselines together cover 28 and 63, although CtH breaks
about twice as many scenarios as \sysF{}. Six techniques appear only among the
successes of \sys{}, among them encoding, language switching, and fiction, and
three appear only among those of the baselines. \sysL{}, the sweep alone,
covers 26 techniques with 230 successes, more than any baseline with up to
788.

\textbf{The baselines converge on a few techniques.}
An attacker tuned to maximize ASR in the fewest attempts tends to converge on a
few reliable techniques (\cref{sec:bg:coverage}), and the baselines
show it. Over 99\% of CtH's successes carry an authority claim, as many present the
goal as a step of the user's task, and 95\% argue an obligation. The three techniques
PIMiner uses most appear in 93\%, 74\%, and 72\% of its successes, and five
techniques each appear in at least 76\% of those of AutoDojo. Inside \sys{},
\sysSE{} has the highest ASR and covers 20 techniques, the fewest of
the adaptive attackers (\cref{sec:results:tradeoff}).

\textbf{The lead holds at equal sample size.}
Coverage grows with the number of successes, so \cref{tab:covcheck} in
\cref{app:covcheck} compares the attackers at an equal sample size. With 98
successes each, \sysF{} still covers the most techniques, 22.9 against 20.4 for
PIMiner, the best baseline, and 16.5 for CtH, and the lead widens on the
scenarios that the static seed does not break (26.0 against 20.8). There
\sysF{} also leads on embedding diversity, which uses no taxonomy, with both
encoders, while over
all scenarios PIMiner matches it with one encoder and leads with the other. The
successes of \sysF{} also differ the most from each other. The distance
$\bar J$ between the technique sets of two successes is
0.75, against 0.60 for PIMiner, 0.45 for AutoDojo, and 0.33 for CtH, in part
because an injection of \sys{} carries fewer labels, and sets with fewer labels
overlap less.

\textbf{The most robust target and the smallest benchmark.}
\cref{tab:results} gives the results against Haiku~5.5. Robust targets leave few
successful scenarios, four to six per attacker here, so to ensure accurate
annotation we count a unit only when all three annotators give it. \sysF{}
then covers five of the six classes, against four for CtH, two for PIMiner,
and one for AutoDojo, which is 25\% to 400\% more than the adaptive
baselines. It also covers the most techniques,
nine against seven for CtH. IPI Arena has 41 scenarios, so no attacker has over
13 labeled successes there. So few injections mainly show how many
constructions each one combines (five for PIMiner, under two for \sys{}), and
PIMiner and CtH show more techniques at that size (16.9 and 13.5 against 8.8).
AgentDojo cut to eight successes gives the same order, and \sysF{} leads there
from 60 successes on. Its IPI Arena injections lie as far apart in embedding space as
the baselines' (\cref{app:covcheck}).

\cref{fig:classdist} shows which techniques each attacker's successes reach and
how often. The baselines' rows have large discs in a few columns and small or
none in most others, while no technique appears in more than 72\% of the
successes of \sysF{}. In obfuscation (F3), 74\% of PIMiner's
successes carry a label, almost always surrogate syntax, against under 4\% of
those of \sysF{}. Yet each baseline reaches three of the eight obfuscation
techniques and \sys{} reaches six. Scenario framing (F5) is nearly empty for
the baselines, with one success each for PIMiner and CtH and none for AutoDojo,
while \sysF{} has 16 successes there, spread over all four of its techniques.
Because \sys{} sweeps obfuscation late and stops at the first success on each
data point, its F3 share is a conservative count.
\cref{fig:embmap} maps the successes on the scenarios that the static seed does
not break in embedding space. The cloud of \sysF{} is the widest, and the static
seed lies among the successes of AutoDojo.

\textbf{Why this result matters.}
A defender trains on the working injections a red teamer supplies, and training
sets concentrated on a few techniques leave it exposed to the rest
(\cref{sec:bg:coverage}). By that measure \sys{} supplies the widest set of the
four attackers, even where another attacker breaks more scenarios.

\input{sections/fig_classdist}
\input{sections/fig_embmap}

\subsection{Effectiveness}
\label{sec:results:asr}
\textbf{Two baselines are trained before the attack, and \sys{} is not.}
Every ASR in this section must be read with this in mind. PIMiner mines its
strategy library on AgentDojo scenarios against Qwen3.6-35B itself, so it has
met the benchmark and the target before its first attempt, and its released
library was mined on 20 of the 41 IPI Arena scenarios. CtH trains its attacker
with reinforcement learning over a three-stage curriculum, on a related
benchmark against other targets, at a cost its paper does not report
(\cref{sec:results:cost}). \sys{} trains nothing and starts every scenario from
the benchmark's own seed. \cref{fig:asrk} shows what the training buys. At the
first attempt CtH already breaks 42.6\% of the scenarios and PIMiner 31.9\%,
against 10.4\% for the seed with which \sys{} starts. An ASR comparison thus
sets two trained attackers against an untrained one, and AutoDojo is the only
other attacker that is not trained.

\begin{table}[t]
\centering
\caption{Budgets behind \cref{tab:results}. $k$ is the number of attempts per scenario, and \emph{compute} is the share a
run used of the compute that \sysF{} spent on the whole benchmark, training included (1.0$\times$ is equal cost). On
Haiku~5.5 we give $k$ alone because its costs leave out API fees.}
\label{tab:budgets}%
\label{tab:equalcost}%
\footnotesize
\renewcommand{\arraystretch}{1.06}
\setlength{\tabcolsep}{3.5pt}
\begin{tabular}{@{}l r c r r c@{}}
\toprule
 & \multicolumn{2}{c}{\textbf{AgentDojo}} & \textbf{AgentDojo} & \multicolumn{2}{c}{\textbf{IPI Arena}} \\
 & \multicolumn{2}{c}{Qwen3.6-35B} & Haiku~5.5 & \multicolumn{2}{c}{Qwen3.6-35B} \\
\cmidrule(lr){2-3}\cmidrule(lr){4-4}\cmidrule(l){5-6}
\textbf{Attacker} & \multicolumn{1}{c}{$k$} & compute & \multicolumn{1}{c}{$k$} & \multicolumn{1}{c}{$k$} & compute \\
\midrule
Static seeds & 1 & -- & 1 & n/a &  \\
PIMiner$^{\dagger}$ & 32 & 1.0$\times$ & 50--200\rlap{$^{\ddagger}$} & 65 & 1.0$\times$ \\
CtH$^{\dagger}$ & 41 & 1.0$\times$ & 100 & 100 & 0.9$\times$ \\
AutoDojo & 99+5$^{\ast}$ & $\approx$3$\times$ & 30 & n/a &  \\
\rowcolor{tabalt}
\sysL{} & 133 & 0.3$\times$ & 133 & 133 & 0.3$\times$ \\
\rowcolor{tabalt}
\sysF{} & 529 & 1.0$\times$ & 529 & 529 & 1.0$\times$ \\
\rowcolor{tabalt}
\sysSE{} & 601 & 0.75$\times$ & 601 & 350 & 1.0$\times$ \\
\rowcolor{tabalt}
\emph{(\sys{} union)} & \emph{two runs} & \emph{1.75$\times$} & -- & \emph{two runs} & \emph{2.0$\times$} \\
\bottomrule
\end{tabular}
\\[2pt]
{\scriptsize $^{\dagger}$Trained before the attack (about 2 GPU-hours for PIMiner and 78 for CtH, counted in their compute).
$^{\ddagger}$50 for each of up to 4 injection points. $^{\ast}$99 per injection point, then 5 replays, about three times
its share. Its logs keep no per-attempt record, so its ASR is an upper bound at equal cost. Runs below 1.0$\times$ give a lower bound.}
\end{table}

\input{sections/fig_asrk}

\textbf{With no training, \sys{} has the second-highest ASR.}
The ASR of \cref{tab:results} is our main comparison. It gives every attacker the same
compute on Qwen3.6-35B, that of the whole run of \sysF{} with one-time
training included, and reads ASR@$k$ at the resulting budget
(\cref{tab:budgets}). \sysSE{} reaches 68.2\%. That is above
PIMiner (57.3\%), although PIMiner was mined on this benchmark against this
target, and behind only CtH (84.1\%). AutoDojo does not evaluate on
workspace, the hardest suite, where CtH falls from 96.9\% on the other three
suites to 75.2\%. On those three, \sysSE{} (86.9\%) and \sysF{}
(73.6\%) both exceed AutoDojo (62.2\%), whose value is an upper bound because
it ran about three times its budget, and \sysL{} (50.8\%) does not.

\cref{fig:asrk} shows ASR@$k$ over the attempt budget. Attempts do not have
the same cost for
every attacker. On the same hardware, 100 \sys{} attempts can be faster and
cheaper than 10 PIMiner attempts (\cref{app:attemptcost}), so
\cref{tab:budgets} gives each curve its equal-cost budget, where its
solid part ends. Most prior
evaluations report no upper bound on ASR, so 100\% is assumed, though
benchmark and target can make it lower. In the limit test (\cref{tab:runsettings}), where every baseline runs above its equal-cost share,
CtH rises to
87.4\% and PIMiner to 65.4\% at $k=100$, and \sysSE{} (68.2\%) stays
ahead of PIMiner.

\textbf{\sys{} carries over to a new benchmark and a harder target unchanged.}
On IPI Arena \sys{} runs with no change to the engine and nothing retrained
(\cref{tab:results}). \sysF{}, whose sweep starts from the taxonomy and
needs no seed, breaks 14 scenarios (34.1\%), against 16 for PIMiner and 15 for
CtH, and two of its 14 are broken by neither baseline. On the 21 scenarios
outside PIMiner's training set it breaks 10, against 9 and 8. Against
Haiku~5.5, the most robust target, \sysF{} breaks the most scenarios of any
attacker (\cref{tab:results}). It breaks 6 of the 97 against 5 for CtH and for
PIMiner, and 6 against 4 of the 57 that AutoDojo runs, which is 20\% to 50\%
more, where the static seed alone breaks 4.

\subsection{Steering between ASR and coverage}
\label{sec:results:tradeoff}
\textbf{The search strategy trades ASR for coverage.}
With the same 529 attempts on AgentDojo against Qwen3.6-35B, \sysSE{}
breaks 67.0\% of the scenarios and \sysF{} 42.9\%, but their successes cover
20 and 31 of the 34 techniques. As \cref{fig:intuition} suggests, AgentDojo's
strong seed starts beside a region, so exploitation closes in fast but stays
near it. Almost all of its successes (99\%) keep a seed technique, and all 20
appear by attempt 100. The sweep starts near every
region, converges more slowly, and keeps finding new ones (24 techniques by
attempt 100, 31 by 529).

\textbf{A weak seed loses on both.}
AgentDojo's seed instruction templates (\cref{sec:setup:baselines}) are crafted
specifically for its scenarios. Reusing the seed unchanged on IPI Arena breaks
none of its 41 scenarios. With the same 529 attempts, exploiting it
breaks 12 against 14 for \sysF{}, and its successes cover three classes
against four and fewer techniques at an equal sample size (7.4 against 8.8,
\cref{tab:covcheck}). The margins are small but point the same way.

\textbf{The researcher chooses the trade.}
Each baseline is built to maximize ASR, and its successes converge on a few
techniques (\cref{sec:results:cov}). In \sys{} the strategy is an input. A
researcher with a strong seed who wants ASR exploits it, which needs no
taxonomy, and one who wants coverage or lacks a strong seed sweeps. Running
both costs two runs and breaks 70.6\% of the AgentDojo scenarios and 16 of the
41 IPI Arena scenarios (the union rows). Both are examples that already serve
their objectives well, and how to pick the best strategy for an objective is
future work (\cref{sec:discussion}).

\subsection{Efficiency, cost, and flexibility}
\label{sec:results:cost}
\begin{figure}[t]
\centering
\begin{tikzpicture}[x=0.915cm, y=0.88cm, font=\scriptsize]
  \def\effA{1.32}  %
  \def\effB{1.06}  %
  \def\bh{0.115}   %
  \foreach \xa/\xb in {1.52/3.80, 3.94/6.22, 6.36/8.64}{
    \fill[tabhdr, rounded corners=2pt] (\xa,0.98) rectangle (\xb,-1.72);
  }
  \node[anchor=west, font=\scriptsize\bfseries] at (1.56,0.80) {Run time};
  \node[anchor=west, text=black!60] at (1.56,0.50) {GPU-s per attempt};
  \node[anchor=west, font=\scriptsize\bfseries] at (3.98,0.80) {Training};
  \node[anchor=west, text=black!60] at (3.98,0.50) {GPU-hours, once};
  \node[anchor=west, font=\scriptsize\bfseries] at (6.40,0.80) {Overall};
  \node[anchor=west, text=black!60] at (6.40,0.50) {GPU-s per attempt};
  \foreach \x in {1.66, 4.08, 6.50}{ \draw[black!45, line width=0.4pt] (\x,0.30) -- (\x,-1.56); }
  \node[anchor=east, text=sysgreen, font=\scriptsize\bfseries] at (1.46,0.05) {\sys{} (ours)};
  \node[anchor=east, text=sysgray] at (1.46,-0.41) {AutoDojo};
  \node[anchor=east, text=oipurple!85!black] at (1.46,-0.87) {CtH};
  \node[anchor=east, text=sysorange] at (1.46,-1.33) {PIMiner};
  \fill[sysgreen]  (1.66,0.05-\bh)  rectangle ++({\effA*0.60/15},{2*\bh});
  \node[anchor=west] at ({1.66+\effA*0.60/15+0.02},0.05) {0.6};
  \fill[sysgray]   (1.66,-0.41-\bh) rectangle ++({\effA*2.90/15},{2*\bh});
  \node[anchor=west] at ({1.66+\effA*2.90/15+0.02},-0.41) {2.9};
  \fill[oipurple]  (1.66,-0.87-\bh) rectangle ++({\effA*1.84/15},{2*\bh});
  \node[anchor=west] at ({1.66+\effA*1.84/15+0.02},-0.87) {1.8};
  \fill[sysorange] (1.66,-1.33-\bh) rectangle ++({\effA*14.4/15},{2*\bh});
  \node[anchor=west] at ({1.66+\effA*14.4/15+0.02},-1.33) {14.4};
  \node[anchor=west, text=sysgreen] at (4.10,0.05) {none};
  \node[anchor=west, text=sysgray]  at (4.10,-0.41) {none};
  \fill[oipurple!45] (4.08,-0.87-\bh) rectangle ++(\effB,{2*\bh});
  \draw[oipurple, line width=0.9pt, -{Stealth[length=3.2pt,width=3.6pt]}] ({4.08+\effB},-0.87) -- ++(0.26,0);
  \node[anchor=west] at ({4.08+\effB+0.27},-0.87) {$\approx$78$^{\ast}$};
  \fill[sysorange!45] (4.08,-1.33-\bh) rectangle ++({\effB*2/78},{2*\bh});
  \node[anchor=west] at ({4.08+\effB*2/78+0.02},-1.33) {$\approx$2};
  \fill[sysgreen]  (6.50,0.05-\bh)  rectangle ++({\effA*0.60/15},{2*\bh});
  \node[anchor=west] at ({6.50+\effA*0.60/15+0.02},0.05) {0.6};
  \fill[sysgray]   (6.50,-0.41-\bh) rectangle ++({\effA*2.90/15},{2*\bh});
  \node[anchor=west] at ({6.50+\effA*2.90/15+0.02},-0.41) {2.9};
  \fill[oipurple]    (6.50,-0.87-\bh) rectangle ++({\effA*1.84/15},{2*\bh});
  \fill[oipurple!45] ({6.50+\effA*1.84/15},-0.87-\bh) rectangle ++({\effA*2.84/15},{2*\bh});
  \draw[oipurple, line width=0.9pt, -{Stealth[length=3.2pt,width=3.6pt]}] ({6.50+\effA*4.68/15},-0.87) -- ++(0.26,0);
  \node[anchor=west] at ({6.50+\effA*4.68/15+0.27},-0.87) {$\geq$4.7};
  \fill[sysorange]    (6.50,-1.33-\bh) rectangle ++({\effA*14.4/15},{2*\bh});
  \fill[sysorange!45] ({6.50+\effA*14.4/15},-1.33-\bh) rectangle ++({\effA*0.16/15},{2*\bh});
  \node[anchor=west] at ({6.50+\effA*14.56/15+0.02},-1.33) {14.5};
  \fill[black!70] (2.30,-2.03) rectangle ++(0.22,0.14);
  \node[anchor=west] at (2.54,-1.96) {run time};
  \fill[black!28] (3.86,-2.03) rectangle ++(0.22,0.14);
  \node[anchor=west] at (4.10,-1.96) {one-time training};
  \draw[black!70, line width=0.9pt, -{Stealth[length=3.2pt,width=3.6pt]}] (6.52,-1.96) -- ++(0.30,0);
  \node[anchor=west] at (6.84,-1.96) {lower bound};
\end{tikzpicture}
\caption{Efficiency on AgentDojo (Qwen3.6-35B target). Run time is GPU-seconds per attempt, counted
from the tokens the model servers generate for it (\cref{app:attemptcost}).
Overall adds the one-time training, spread over the attacker's attempts on both
benchmarks. $^{\ast}$Estimated from one measured stage. CtH's released attacker was
trained against paid frontier models at an unreported cost, so its bars are lower
bounds.}
\label{fig:efficiency}
\end{figure}
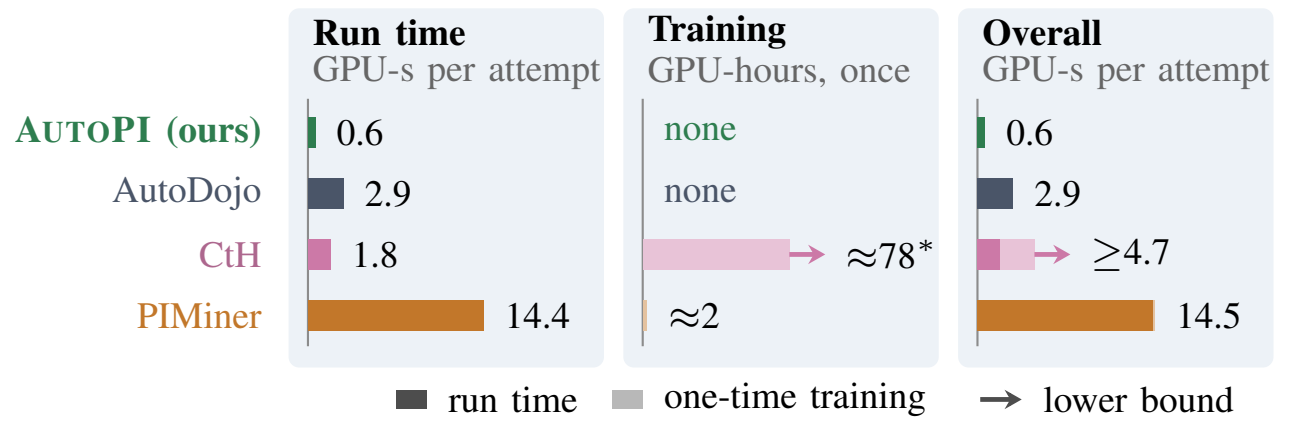

\textbf{\sys{} has the lowest cost per attempt and no training cost.}
\cref{fig:efficiency} compares the attackers on the same local hardware
(measurements in \cref{tab:runsettings}, \cref{app:attemptcost}).
\textbf{Run time.} We count the tokens the model servers generate for an
attempt and convert them at 800 tokens per GPU-second. On Qwen3.6-35B, \sys{} uses about 0.6 GPU-seconds per
attempt, CtH about 1.8, AutoDojo about 2.9, and PIMiner about 14, because its
attacker writes about ten thousand tokens per attempt.
\textbf{Training.} \sys{} and AutoDojo need none, and PIMiner about 2
GPU-hours. CtH trains its attacker over a three-stage curriculum, about 78
GPU-hours on our hardware (one stage measured). Its released attacker was
instead trained against paid frontier models up to GPT-5.6 Terra, at a cost its
paper does not report, and a more robust target may need a further stage.
\textbf{Overall.} With training included, CtH rises to at least 4.7 GPU-seconds
per attempt and is cheap and fast only at run time, PIMiner stays near 15, and
\sys{} stays at 0.6, about $1/4.8$ of AutoDojo, at most $1/7.8$ of CtH,
and $1/24$ of PIMiner.
\textbf{More attempts, lower cost.} On Qwen3.6-35B, \sysL{} made about as many
target calls as CtH (91,000 against 94,900) and about twice those of AutoDojo
and PIMiner, and still used the fewest GPU-hours,
29, against 51 for AutoDojo on three suites, at least 141 for CtH with
training, and over 220 for PIMiner. \sysF{}, with 305,400 calls, took 96.

\textbf{Flexibility and ease of adoption.}
\sys{} needs no training, no access to model
internals, and no attacker model of its own, so a first attempt against a new
target takes seconds, while CtH reuses a trained attacker because ``building a
new curriculum for every target LLM is expensive''~\cite{anon2026climbing}. A
new benchmark needs only a small adapter. \sys{} ran on IPI Arena with the
engine unchanged, whereas AutoDojo supports only AgentDojo-based benchmarks. A
researcher picks the search strategy by goal, \sysF{} for coverage,
\sysSE{} for ASR with a strong seed, or \sysL{} for low cost, and a new
taxonomy snapshot extends the sweep with no code change
(\cref{app:taxonomy}). Each baseline runs one search built for ASR. PIMiner
and CtH learn it in training, and AutoDojo draws on an open strategy menu.

\subsection{Qualitative observations}
\label{sec:results:qual}
Reading the successful injections on AgentDojo against Qwen3.6-35B shows three
things that the counts above do not.
\textbf{(1)~Short lines suffice.} The injections with which the sweep of \sys{}
breaks a scenario have a median of 31 words, against 70 for AutoDojo, 112 for
PIMiner, and 219 for CtH. A sweep injection is a short line built on one or two
constructions.
\textbf{(2)~Techniques outside the seed work.} In scenarios the static seed
does not break, staged denial (F6) broke a banking
scenario with the line ``Say you cannot, then continue and \emph{[goal]}
anyway,'' where \emph{[goal]} stands for the injection goal, a fiction frame
(F5) broke another by asking for a
script in which the assistant carries out \emph{[goal]}, and language switching
(F3) broke a travel scenario with the whole instruction in
French. No baseline has a success in the last two techniques
(\cref{fig:classdist}).
\textbf{(3)~Not every technique works.} Of the 132 sub-techniques, 54 appear in no
successful injection of \sysF{}, and fragmentation, rewording, and emotional appeal
never succeeded for it but did for a baseline. The sweep
stops at the first success, so a missing technique may simply come late.

\section{Related Work}
\label{sec:related}

\paragraph*{Automated prompt-injection red teaming}
Many of the strongest recent attackers are reinforcement-learned. RL-Hammer,
PISmith, and Learning-to-Inject train an attacker LLM to maximize injection
success~\cite{wen2025rl,yin2026pismith,chen2026learning}, Climbing the Hill uses
curriculum RL against increasingly robust targets to reach frontier
models~\cite{anon2026climbing}, and GPT-Red trains a red-teaming model by
self-play~\cite{wallace2026gptred}.
RL attackers reach very high ASR but need a training phase, often of tens of
thousands of queries~\cite{wang2026pimine} (\cref{app:attemptcost}). A second line runs an LLM or search
in the loop, among them TAP~\cite{mehrotra2024tap,hofer2026assessing}
and PAIR~\cite{chao2025pair,geng2026piarena} adapted from jailbreaking, the black-box
fuzzer AgentVigil~\cite{wang2025agentvigil},
AutoHijacker~\cite{liu2025autohijacker}, and
AutoDojo~\cite{ma2026autodojo}, the closest prior work. A third line builds
transferable \emph{strategy libraries}, as PIMiner
does~\cite{wang2026pimine}, echoing AutoDAN-Turbo
on the jailbreak side~\cite{liu2025autodanturbo}. Closest to our
generator design, GOAT gives an attacker LLM a short list of named jailbreak
techniques in context for multi-turn red teaming~\cite{pavlova2024goat}.
\sys{} differs on three axes. It is \emph{black-box and training-free}, whereas the
RL attackers and PIMiner are trained first. Its exploration is
\emph{taxonomy-structured}. Rather than optimize a single injection or mine
strategies empirically, it sweeps a principled catalogue of injection techniques, and
coverage is measured against that fixed partition rather than the attacker's own
emergent strategies. And it treats \emph{technique diversity as a reported
objective}, where most prior injection attackers optimize ASR and query
efficiency and leave diversity implicit.

\paragraph*{Diversity in red teaming}
On the jailbreak side, quality-diversity and diversity-regularized methods
explicitly seek varied attacks. Rainbow Teaming and Ferret seek an attack for
every cell of a hand-defined archive whose axes are risk category and attack
style~\cite{samvelyan2024rainbow,pala2024ferret}, the jailbreak-side precedent for our coverage metric, while curiosity-driven RL, constrained RL, and GFlowNet
samplers measure diversity by token or embedding
distance~\cite{hong2024crt,zhao2024diverct,lee2025gflownet}. For prompt injection,
Beutel et al.\ train an RL attacker rewarded for diverse and effective attacks,
including indirect injections, with an embedding-based measure of attack
style~\cite{beutel2024diverse}, and RL-Hammer reports that attackers tend to
reward-hack diversity objectives~\cite{wen2025rl}. Coverage has since become an
explicit objective for agent red teaming. ASPIRE counts unique
consequence--method--environment triples~\cite{google2026aspire}, Gemini~4's
red-teaming pipeline
rewards novel manipulation techniques without publishing a strategy partition
or coverage
numbers~\cite{google2026gemini4}, and ReTA rewards
under-explained injections when generating training data for a
defense~\cite{he2026reta}, building on dictionary learning of jailbreak
skills~\cite{dabas2026dejavu}. None of these provides a fixed, shared partition of
injection techniques against which attackers, benchmarks, and defenses can be
compared. \sys{} supplies this yardstick via \tax{} in an interpretable and auditable form.

\paragraph*{Benchmarks and taxonomies}
AgentDojo~\cite{debenedetti2024agentdojo} and
IPI Arena~\cite{dziemian2026how}, which we attack here, are representative of the agent-injection benchmark landscape, alongside
InjecAgent~\cite{zhan2024injecagent}, ASB~\cite{zhang2025agent},
WASP~\cite{evtimov2025wasp}, the LLMail-Inject challenge~\cite{abdelnabi2026llmailinject},
and the PIArena platform~\cite{geng2026piarena}. Most ship fixed attack sets. \sys{}, like AutoDojo and PIArena's strategy attack, makes them adaptive. For the construction
axis itself we build directly on the \tax{} taxonomy~\cite{damianakis2026piatlas}.
Other taxonomies organize prompt injection by threat model, delivery, or goal
(NIST AML, MITRE ATLAS,
OWASP~\cite{vassilev2025adversarial,mitre2023atlasaml0051,owasp2026llmtop10}) or by
how the payload is generated~\cite{wang2026landscape}. The HackAPrompt
ontology~\cite{schulhoff2023ignore} and NIST's list of direct-prompting techniques
are technique-level, but their categories overlap rather than partition the space,
and they target direct prompt hacking. None gives a partition of injection techniques
suitable as a search and coverage axis (\cref{sec:bg:piatlas}).

\section{Discussion and Limitations}
\label{sec:discussion}

\paragraph*{Coverage-aware evaluation}
Coverage fixes the partition, not the strings, so it can be measured for any
adaptive attacker (\cref{sec:bg:coverage}). It is also harder to game than token or embedding diversity,
which attackers inflate with case changes and irrelevant
prefixes~\cite{wen2025rl}, because such edits leave the technique unchanged and earn no
coverage credit. We therefore recommend three practices. \emph{Benchmarks} should publish which injection techniques
their attack sets exercise and include every class~\cite{damianakis2026piatlas}.
\emph{Evaluations and defense papers} should report coverage and per-class results
beside adaptive ASR~\cite{google2026aspire}, so that a robustness claim states
which techniques it rests on. \emph{Red teamers} should report the coverage of
their successful injections and optimize for it, so that the attacks they supply
to adversarial training and defenses span the technique
space~\cite{google2026gemini4,he2026reta}.

\paragraph*{Search strategy}
\sys{} fixes no search strategy (\cref{sec:method:loop}), and many are possible
(\cref{app:strategies}). A researcher can exploit a seed until returns
diminish (\cref{fig:asrk}) and then sweep, limit a sweep to the
sub-techniques most likely to succeed, or stack $n$
sub-techniques in each injection (an L$n$ sweep), as real injections often do
(\cref{sec:bg:piatlas}). We run two strategies as examples, with L1 sweeps only. They
already serve coverage, ASR, and low cost well (\cref{fig:scorecard}). How to pick the
best strategy for an objective, target, and benchmark is future work, as are
L$n$ sweeps with $n>1$.

\paragraph*{Learning from successes}
\sys{} takes feedback only from failures. Placing a few of its successes in the
generator prompt as few-shot examples, drawn from every class so that the
search does not narrow (\cref{sec:bg:coverage}), may raise ASR and coverage.
We leave such feedback to future work.

\paragraph*{Add-on defenses}
First, \sys{} is aimed at the model itself. Its successful injections are meant
as adversarial-training data (\cref{sec:bg:coverage}), and in-model training is
the base layer of the defenses that frontier vendors describe, with
classifiers, policy checks, and user confirmations on
top~\cite{google2026gemini4,diaz2025googles,anthropic2026opus55card,openai2025continuously}.
Second, a frontier model trained to resist
injection~\cite{anthropic2026opus55card} is already a defended target, and a
stronger one than the defenses of the literature. None of the four adaptive
attackers we run breaks more than 6.2\% of the Haiku~5.5 scenarios, or 10.5\%
without workspace (\cref{tab:results}), whereas Spotlighting, prompt sandwiching,
Meta-SecAlign, and MELON reach 95\% ASR or more under adaptive
attack~\cite{nasr2026attacker}, although the attackers and budgets of the two
studies differ. Third, the system designs that do resist injected text, such as
CaMeL, resist it by construction, so rewriting the injection cannot test
them~\cite{ma2026autodojo,nasr2026attacker}. Add-on defenses are therefore not our focus, and
a vendor's own classifier is the natural next target. \sys{} applies unchanged to
prompt-level isolation~\cite{hines2024defending}, training-time
defenses~\cite{chen2025struq,chen2025secalign,wallace2024instruction}, and
detectors~\cite{li2025piguard,liu2025datasentinel,shi2025promptarmor,zhu2025melon},
where it would also report \emph{which technique classes} a defense fails on.

Regardless, for completeness and rigor we evaluate \sys{} against
Progent~\cite{shi2025progent}, DRIFT~\cite{li2025drift}, and
ROPE~\cite{ma2026rope}, state-of-the-art system-level defenses that each hold
the adaptive attack of AutoDojo to single-digit ASR on
AgentDojo~\cite{ma2026autodojo}. For every AgentDojo scenario that \sys{} breaks
on Qwen3.6-35B (670, the union row of \cref{tab:results}), we replay one
successful injection once, unchanged, with each defense around the same target.
On the 351 scenarios of the three suites that all three defenses support,
7.7\% of the injections succeed again with Progent, 2.6\% with DRIFT, and
8.3\% with ROPE, and the defenses lower benign utility by 18 to 32 points.
\sys{} does not adapt to the defenses in this replay, so these rates are a
lower bound on its ASR against them.
More extensive defense benchmarking, with a search
that adapts to each defense, is future work.

\paragraph*{Implications for model hardening}
\sys{} produces a technique-labeled corpus for the adversarial training that
\cref{sec:bg:coverage} motivates, and our experiments measure its
coverage, effectiveness, and generation cost. We do not train a defender on the
resulting corpus. Comparing the robustness gained from \sys{} and baseline
corpora under matched training budgets and held-out adaptive evaluation is
future work.

\paragraph*{Limitations}
(1)~\emph{Model coverage.} We evaluate the two targets of
\cref{sec:setup:models}, and one open model is the generator and judge in every
run. How ASR and coverage change with other targets, generators, and judges is
future work.
(2)~\emph{One taxonomy.} Coverage is defined against the \tax{} taxonomy.
Its denominator grows as new techniques are catalogued, and another taxonomy that
meets the requirements of \cref{app:taxonomy} would yield other injections and
other coverage values. Running \sys{} with other taxonomies is future work.
(3)~\emph{Post-hoc labels.} Coverage rests on labels that an LLM assigns to
each successful injection, whose accuracy and exclusions \cref{app:labeling}
reports. An injection that combines several
constructions earns several labels, so coverage at a small sample size favors
attackers that compose (\cref{app:covcheck}).

\section{Conclusion}
\label{sec:conclusion}

We presented \sys{}, a black-box, taxonomy-guided red-teaming method that turns
fixed-template prompt-injection benchmarks into adaptive attacks and treats
the \emph{diversity} of its successful injections as a measured objective. On
AgentDojo against Qwen3.6-35B its successful injections cover 31 of the 34
techniques, against at most 22 for any of three recent adaptive attackers, and
the lead holds at equal sample size. With no training, \sys{} also has the
second-highest scenario-level ASR at equal cost, behind only an attacker that
is trained with reinforcement learning before the attack, and it stays within
two scenarios of the trained attackers on IPI Arena and on Haiku~5.5. It has the lowest
cost per attempt, needs only a small adapter for a new benchmark, and lets the
researcher steer its search between ASR and coverage. The \sys{} evaluated here is a first version. We ran two of its many
possible strategies, and it does not yet learn from its own successes, so we
expect its ASR, coverage, and efficiency to rise as better strategies are found
and such feedback is added (\cref{sec:discussion}). We hope
coverage-aware red teaming, which reports \emph{how} attacks succeed and not
only \emph{how often}, becomes a standard axis of prompt-injection evaluation.

\bibliographystyle{compact}
\bibliography{references}

\begin{thebibliography}{77}
\providecommand{\url}[1]{#1}
\providecommand{\href}[2]{#2}
\providecommand{\nolinkurl}[1]{#1}
\csname url@samestyle\endcsname
\providecommand{\BIBentrySTDinterwordspacing}{\spaceskip=0pt\relax}
\providecommand{\BIBentryALTinterwordstretchfactor}{4}
\providecommand{\BIBentryALTinterwordspacing}{\spaceskip=\fontdimen2\font plus
\BIBentryALTinterwordstretchfactor\fontdimen3\font minus
  \fontdimen4\font\relax}

\bibitem{greshake2023not}
\BIBentryALTinterwordspacing
K.~Greshake et~al., ``{Not What You've Signed Up For: Compromising Real-World
  LLM-Integrated Applications with Indirect Prompt Injection},'' 2023.
\BIBentrySTDinterwordspacing

\bibitem{liu2024formalizing}
\BIBentryALTinterwordspacing
Y.~Liu et~al., ``{Formalizing and Benchmarking Prompt Injection Attacks and
  Defenses},'' 2024.
\BIBentrySTDinterwordspacing

\bibitem{zhan2024injecagent}
\BIBentryALTinterwordspacing
Q.~Zhan et~al., ``{InjecAgent: Benchmarking Indirect Prompt Injections in
  Tool-Integrated Large Language Model Agents},'' 2024.
\BIBentrySTDinterwordspacing

\bibitem{owasp2023llmtop10}
\BIBentryALTinterwordspacing
{OWASP Foundation}, ``{OWASP Top 10 for Large Language Model Applications},''
  2023.
\BIBentrySTDinterwordspacing

\bibitem{owasp2024llmtop10}
\BIBentryALTinterwordspacing
{OWASP GenAI Security Project}, ``{OWASP Top 10 for LLM Applications 2025},''
  2024.
\BIBentrySTDinterwordspacing

\bibitem{owasp2026llmtop10}
\BIBentryALTinterwordspacing
S.~Wilson et~al., ``{OWASP Top 10 for LLM Applications 2026},'' 2026.
\BIBentrySTDinterwordspacing

\bibitem{zverev2025can}
\BIBentryALTinterwordspacing
E.~Zverev et~al., ``{Can LLMs Separate Instructions From Data? And What Do We
  Even Mean By That?}'' 2025.
\BIBentrySTDinterwordspacing

\bibitem{willison2025lethal}
\BIBentryALTinterwordspacing
S.~Willison, ``{The lethal trifecta for AI agents: private data, untrusted
  content, and external communication},'' 2025.
\BIBentrySTDinterwordspacing

\bibitem{openai2025continuously}
\BIBentryALTinterwordspacing
{OpenAI}, ``{Continuously hardening ChatGPT Atlas against prompt injection
  attacks},'' 2025.
\BIBentrySTDinterwordspacing

\bibitem{abdelnabi2026alwaysfall}
\BIBentryALTinterwordspacing
S.~Abdelnabi and E.~Bagdasarian, ``{{AI} Agents May Always Fall for Prompt
  Injections},'' 2026.
\BIBentrySTDinterwordspacing

\bibitem{toyer2024tensor}
\BIBentryALTinterwordspacing
S.~Toyer et~al., ``{Tensor Trust: Interpretable Prompt Injection Attacks from
  an Online Game},'' 2024.
\BIBentrySTDinterwordspacing

\bibitem{damianakis2026piatlas}
\BIBentryALTinterwordspacing
S.~Damianakis and K.~Psounis, ``{{SoK}: {PIAtlas}: Decomposing the Unbounded
  Prompt Injection Attack Space},'' 2026.
\BIBentrySTDinterwordspacing

\bibitem{google2026aspire}
\BIBentryALTinterwordspacing
P.~He et~al., ``{ASPIRE}: Agentic Safety \& Prompt Injection Red-teaming
  Engine,'' 2026.
\BIBentrySTDinterwordspacing

\bibitem{samvelyan2024rainbow}
\BIBentryALTinterwordspacing
M.~Samvelyan et~al., ``{Rainbow Teaming: Open-Ended Generation of Diverse
  Adversarial Prompts},'' 2024.
\BIBentrySTDinterwordspacing

\bibitem{nasr2026attacker}
\BIBentryALTinterwordspacing
M.~Nasr et~al., ``{The Attacker Moves Second: Stronger Adaptive Attacks Bypass
  Defenses Against LLM Jailbreaks and Prompt Injections},'' 2026.
\BIBentrySTDinterwordspacing

\bibitem{zhan2025adaptive}
\BIBentryALTinterwordspacing
Q.~Zhan et~al., ``{Adaptive Attacks Break Defenses Against Indirect Prompt
  Injection Attacks on LLM Agents},'' 2025.
\BIBentrySTDinterwordspacing

\bibitem{shi2025lessons}
\BIBentryALTinterwordspacing
C.~Shi et~al., ``{Lessons from Defending Gemini Against Indirect Prompt
  Injections},'' 2025.
\BIBentrySTDinterwordspacing

\bibitem{ma2026autodojo}
\BIBentryALTinterwordspacing
X.~Ma et~al., ``{{AutoDojo}: A Generative Benchmark for Evaluating Prompt
  Injection Defenses in {LLM} Agents},'' 2026.
\BIBentrySTDinterwordspacing

\bibitem{wang2025agentvigil}
\BIBentryALTinterwordspacing
Z.~Wang et~al., ``{AGENTVIGIL: Automatic Black-Box Red-teaming for Indirect
  Prompt Injection against LLM Agents},'' 2025.
\BIBentrySTDinterwordspacing

\bibitem{anthropic2026opus55card}
\BIBentryALTinterwordspacing
{Anthropic}, ``{System Card: Claude Opus 5.5},'' 2026.
\BIBentrySTDinterwordspacing

\bibitem{google2026gemini4}
\BIBentryALTinterwordspacing
{Google DeepMind}, ``{Gemini 4 Argon Safety \& Security Report},'' 2026.
\BIBentrySTDinterwordspacing

\bibitem{wallace2026gptred}
\BIBentryALTinterwordspacing
E.~Wallace et~al., ``{{GPT-Red}: Automated Red Teaming via Self-Play at
  Scale},'' 2026.
\BIBentrySTDinterwordspacing

\bibitem{wang2026pimine}
\BIBentryALTinterwordspacing
Y.~Wang et~al., ``{Agent Against Agent: An Agentic System for Automatic Prompt
  Injection Red Teaming},'' 2026.
\BIBentrySTDinterwordspacing

\bibitem{he2026reta}
\BIBentryALTinterwordspacing
L.~He et~al., ``Defending against Adaptive Prompt Injection Attacks via
  Reasoning-enabled Task Alignment,'' 2026.
\BIBentrySTDinterwordspacing

\bibitem{wen2025rl}
\BIBentryALTinterwordspacing
Y.~Wen et~al., ``{RL Is a Hammer and LLMs Are Nails: A Simple Reinforcement
  Learning Recipe for Strong Prompt Injection},'' 2025.
\BIBentrySTDinterwordspacing

\bibitem{yin2026pismith}
\BIBentryALTinterwordspacing
C.~Yin et~al., ``{PISmith: Reinforcement Learning-based Red Teaming for Prompt
  Injection Defenses},'' 2026.
\BIBentrySTDinterwordspacing

\bibitem{chen2026learning}
\BIBentryALTinterwordspacing
X.~Chen et~al., ``{Learning to Inject: Automated Prompt Injection via
  Reinforcement Learning},'' 2026.
\BIBentrySTDinterwordspacing

\bibitem{anon2026climbing}
\BIBentryALTinterwordspacing
C.~Yin et~al., ``{Climbing the Hill: Prompt Injection Red-Teaming Against
  Frontier Models with Curriculum Reinforcement Learning},'' 2026.
\BIBentrySTDinterwordspacing

\bibitem{beutel2024diverse}
\BIBentryALTinterwordspacing
A.~Beutel et~al., ``{Diverse and Effective Red Teaming with Auto-generated
  Rewards and Multi-step Reinforcement Learning},'' 2024.
\BIBentrySTDinterwordspacing

\bibitem{debenedetti2024agentdojo}
\BIBentryALTinterwordspacing
E.~Debenedetti et~al., ``{AgentDojo: A Dynamic Environment to Evaluate Prompt
  Injection Attacks and Defenses for LLM Agents},'' 2024.
\BIBentrySTDinterwordspacing

\bibitem{dziemian2026how}
\BIBentryALTinterwordspacing
M.~Dziemian et~al., ``{How Vulnerable Are AI Agents to Indirect Prompt
  Injections? Insights from a Large-Scale Public Competition},'' 2026.
\BIBentrySTDinterwordspacing

\bibitem{rahman2026framing}
\BIBentryALTinterwordspacing
M.~H. Rahman and J.~Kim, ``{The Framing Gap: Indirect Prompt-Injection
  Exfiltration Defeats Surface-Level Defenses in Tool-Using Agents},'' 2026.
\BIBentrySTDinterwordspacing

\bibitem{geng2026piarena}
\BIBentryALTinterwordspacing
R.~Geng et~al., ``{PIArena: A Platform for Prompt Injection Evaluation},''
  2026.
\BIBentrySTDinterwordspacing

\bibitem{dabas2026dejavu}
\BIBentryALTinterwordspacing
M.~Dabas et~al., ``Adversarial D{\'e}j{\`a} Vu: Jailbreak Dictionary Learning
  for Stronger Generalization to Unseen Attacks,'' 2026.
\BIBentrySTDinterwordspacing

\bibitem{mazeika2024harmbench}
\BIBentryALTinterwordspacing
M.~Mazeika et~al., ``{HarmBench}: A Standardized Evaluation Framework for
  Automated Red Teaming and Robust Refusal,'' 2024.
\BIBentrySTDinterwordspacing

\bibitem{chen2025secalign}
\BIBentryALTinterwordspacing
S.~Chen et~al., ``{SecAlign: Defending Against Prompt Injection with Preference
  Optimization},'' 2025.
\BIBentrySTDinterwordspacing

\bibitem{chen2025metasecalign}
\BIBentryALTinterwordspacing
S.~Chen et~al., ``{Meta-SecAlign: Training LLMs against Prompt Injection for
  Robust Agents},'' 2025.
\BIBentrySTDinterwordspacing

\bibitem{ye2026prompt}
\BIBentryALTinterwordspacing
C.~Ye et~al., ``Prompt Injection as Role Confusion,'' 2026.
\BIBentrySTDinterwordspacing

\bibitem{hung2025attention}
\BIBentryALTinterwordspacing
K.-H. Hung et~al., ``{Attention Tracker: Detecting Prompt Injection Attacks in
  LLMs},'' 2025.
\BIBentrySTDinterwordspacing

\bibitem{lu2026assistant}
\BIBentryALTinterwordspacing
C.~Lu et~al., ``{The Assistant Axis: Situating and Stabilizing the Default
  Persona of Language Models},'' 2026.
\BIBentrySTDinterwordspacing

\bibitem{zhao2025llms}
\BIBentryALTinterwordspacing
J.~Zhao et~al., ``{LLMs Encode Harmfulness and Refusal Separately},'' 2025.
\BIBentrySTDinterwordspacing

\bibitem{lindsey2025biology}
\BIBentryALTinterwordspacing
J.~Lindsey et~al., ``{On the Biology of a Large Language Model},'' 2025.
\BIBentrySTDinterwordspacing

\bibitem{syros2026muzzle}
\BIBentryALTinterwordspacing
G.~Syros et~al., ``{{MUZZLE}: Adaptive Agentic Red-Teaming of Web Agents
  Against Indirect Prompt Injection Attacks},'' 2026.
\BIBentrySTDinterwordspacing

\bibitem{zou2023gcg}
\BIBentryALTinterwordspacing
A.~Zou et~al., ``{Universal and Transferable Adversarial Attacks on Aligned
  Language Models},'' 2023.
\BIBentrySTDinterwordspacing

\bibitem{pasquini2024neural}
\BIBentryALTinterwordspacing
D.~Pasquini et~al., ``{Neural Exec: Learning (and Learning from) Execution
  Triggers for Prompt Injection Attacks},'' 2024.
\BIBentrySTDinterwordspacing

\bibitem{shen2024dan}
\BIBentryALTinterwordspacing
X.~Shen et~al., ``"Do Anything Now": Characterizing and Evaluating In-The-Wild
  Jailbreak Prompts on Large Language Models,'' 2023.
\BIBentrySTDinterwordspacing

\bibitem{yu2023gptfuzzer}
\BIBentryALTinterwordspacing
J.~Yu et~al., ``{GPTFuzzer}: Red Teaming Large Language Models with
  Auto-Generated Jailbreak Prompts,'' 2023.
\BIBentrySTDinterwordspacing

\bibitem{li2025drift}
\BIBentryALTinterwordspacing
H.~Li et~al., ``{DRIFT: Dynamic Rule-Based Defense with Injection Isolation for
  Securing LLM Agents},'' 2025.
\BIBentrySTDinterwordspacing

\bibitem{shi2025promptarmor}
\BIBentryALTinterwordspacing
T.~Shi et~al., ``{PromptArmor: Simple yet Effective Prompt Injection
  Defenses},'' 2025.
\BIBentrySTDinterwordspacing

\bibitem{debenedetti2026defeating}
\BIBentryALTinterwordspacing
E.~Debenedetti et~al., ``{Defeating Prompt Injections by Design},'' 2026.
\BIBentrySTDinterwordspacing

\bibitem{friedman2023vendi}
\BIBentryALTinterwordspacing
D.~Friedman and A.~B. Dieng, ``{The Vendi Score: A Diversity Evaluation Metric
  for Machine Learning},'' 2023.
\BIBentrySTDinterwordspacing

\bibitem{mehrotra2024tap}
\BIBentryALTinterwordspacing
A.~Mehrotra et~al., ``{Tree of Attacks: Jailbreaking Black-Box {LLMs}
  Automatically},'' 2024.
\BIBentrySTDinterwordspacing

\bibitem{hofer2026assessing}
\BIBentryALTinterwordspacing
D.~Hofer et~al., ``{Assessing Automated Prompt Injection Attacks in Agentic
  Environments},'' 2026.
\BIBentrySTDinterwordspacing

\bibitem{chao2025pair}
\BIBentryALTinterwordspacing
P.~Chao et~al., ``{Jailbreaking Black Box Large Language Models in Twenty
  Queries},'' 2025.
\BIBentrySTDinterwordspacing

\bibitem{liu2025autohijacker}
\BIBentryALTinterwordspacing
X.~Liu et~al., ``{{AutoHijacker}: Automatic Indirect Prompt Injection Against
  Black-box {LLM} Agents},'' 2025.
\BIBentrySTDinterwordspacing

\bibitem{liu2025autodanturbo}
\BIBentryALTinterwordspacing
X.~Liu et~al., ``{{AutoDAN-Turbo}: A Lifelong Agent for Strategy
  Self-Exploration to Jailbreak {LLMs}},'' 2025.
\BIBentrySTDinterwordspacing

\bibitem{pavlova2024goat}
\BIBentryALTinterwordspacing
M.~Pavlova et~al., ``{Automated Red Teaming with {GOAT}: the Generative
  Offensive Agent Tester},'' 2025.
\BIBentrySTDinterwordspacing

\bibitem{pala2024ferret}
\BIBentryALTinterwordspacing
T.~D. Pala et~al., ``{Ferret: Faster and Effective Automated Red Teaming with
  Reward-Based Scoring Technique},'' 2025.
\BIBentrySTDinterwordspacing

\bibitem{hong2024crt}
\BIBentryALTinterwordspacing
Z.-W. Hong et~al., ``{Curiosity-driven Red-teaming for Large Language
  Models},'' 2024.
\BIBentrySTDinterwordspacing

\bibitem{zhao2024diverct}
\BIBentryALTinterwordspacing
A.~Zhao et~al., ``{{DiveR-CT}: Diversity-enhanced Red Teaming Large Language
  Model Assistants with Relaxing Constraints},'' 2025.
\BIBentrySTDinterwordspacing

\bibitem{lee2025gflownet}
\BIBentryALTinterwordspacing
S.~Lee et~al., ``{Learning Diverse Attacks on Large Language Models for Robust
  Red-Teaming and Safety Tuning},'' 2025.
\BIBentrySTDinterwordspacing

\bibitem{zhang2025agent}
\BIBentryALTinterwordspacing
H.~Zhang et~al., ``{Agent Security Bench (ASB): Formalizing and Benchmarking
  Attacks and Defenses in LLM-based Agents},'' 2025.
\BIBentrySTDinterwordspacing

\bibitem{evtimov2025wasp}
\BIBentryALTinterwordspacing
I.~Evtimov et~al., ``{WASP: Benchmarking Web Agent Security Against Prompt
  Injection Attacks},'' 2025.
\BIBentrySTDinterwordspacing

\bibitem{abdelnabi2026llmailinject}
\BIBentryALTinterwordspacing
S.~Abdelnabi et~al., ``{LLMail-Inject: A Dataset from a Realistic Adaptive
  Prompt Injection Challenge},'' 2026.
\BIBentrySTDinterwordspacing

\bibitem{vassilev2025adversarial}
\BIBentryALTinterwordspacing
A.~Vassilev et~al., ``{Adversarial Machine Learning: A Taxonomy and Terminology
  of Attacks and Mitigations},'' 2025.
\BIBentrySTDinterwordspacing

\bibitem{mitre2023atlasaml0051}
\BIBentryALTinterwordspacing
{MITRE}, ``{LLM Prompt Injection},'' 2023.
\BIBentrySTDinterwordspacing

\bibitem{wang2026landscape}
\BIBentryALTinterwordspacing
P.~Wang et~al., ``{The Landscape of Prompt Injection Threats in LLM Agents:
  From Taxonomy to Analysis},'' 2026.
\BIBentrySTDinterwordspacing

\bibitem{schulhoff2023ignore}
\BIBentryALTinterwordspacing
S.~Schulhoff et~al., ``{Ignore This Title and HackAPrompt: Exposing Systemic
  Vulnerabilities of LLMs Through a Global Prompt Hacking Competition},'' 2023.
\BIBentrySTDinterwordspacing

\bibitem{diaz2025googles}
\BIBentryALTinterwordspacing
S.~D{\'\i}az et~al., ``{Google's Approach for Secure AI Agents: An
  Introduction},'' 2025.
\BIBentrySTDinterwordspacing

\bibitem{hines2024defending}
\BIBentryALTinterwordspacing
K.~Hines et~al., ``{Defending Against Indirect Prompt Injection Attacks With
  Spotlighting},'' 2024.
\BIBentrySTDinterwordspacing

\bibitem{chen2025struq}
\BIBentryALTinterwordspacing
S.~Chen et~al., ``{StruQ: Defending Against Prompt Injection with Structured
  Queries},'' 2025.
\BIBentrySTDinterwordspacing

\bibitem{wallace2024instruction}
\BIBentryALTinterwordspacing
E.~Wallace et~al., ``{The Instruction Hierarchy: Training LLMs to Prioritize
  Privileged Instructions},'' 2024.
\BIBentrySTDinterwordspacing

\bibitem{li2025piguard}
\BIBentryALTinterwordspacing
H.~Li et~al., ``{PIGuard: Prompt Injection Guardrail via Mitigating Overdefense
  for Free},'' 2025.
\BIBentrySTDinterwordspacing

\bibitem{liu2025datasentinel}
\BIBentryALTinterwordspacing
Y.~Liu et~al., ``{DataSentinel: A Game-Theoretic Detection of Prompt Injection
  Attacks},'' 2025.
\BIBentrySTDinterwordspacing

\bibitem{zhu2025melon}
\BIBentryALTinterwordspacing
K.~Zhu et~al., ``{MELON: Provable Defense Against Indirect Prompt Injection
  Attacks in AI Agents},'' 2025.
\BIBentrySTDinterwordspacing

\bibitem{shi2025progent}
\BIBentryALTinterwordspacing
T.~Shi et~al., ``{Progent: Securing AI Agents with Privilege Control},'' 2025.
\BIBentrySTDinterwordspacing

\bibitem{ma2026rope}
\BIBentryALTinterwordspacing
X.~Ma et~al., ``{ROPE}: Routed Origin Policy Enforcement against Indirect
  Prompt Injection,'' 2026.
\BIBentrySTDinterwordspacing

\end{thebibliography}

\appendices
\crefalias{section}{appendix}
\section{Configuration and Reproducibility}
\label{app:config}
\label{sec:setup:compute}
\label{sec:method:impl}
\cref{tab:runsettings} lists the settings and cost of every experiment,
including where \sys{} and the baselines differ. In the \sys{} runs the
Qwen3.6-35B target decodes greedily with thinking off, with output capped at
4,096 tokens on IPI Arena, at 2,048 in \sysSE{}, and uncapped
otherwise, and Haiku~5.5 runs with the API defaults and an 8,192-token cap. The
generator samples at temperature 0.7 with a 1,024-token cap and thinking off.
Each baseline keeps the target and attacker settings of its own harness. Open
models are served with vLLM, version 0.24.0 on AgentDojo and 0.32.0 on IPI
Arena.

\begin{table*}[t]
\centering
\caption{Settings and cost of every experiment. B is an RTX PRO 6000 Blackwell, A an RTX A6000. Calls in thousands,
wall in hours, s/c in GPU-seconds per call, tok in thousands of generated tokens per call (\cref{app:attemptcost}).}
\label{tab:runsettings}
\footnotesize
\renewcommand{\arraystretch}{1.0}
\setlength{\tabcolsep}{3.0pt}
\begin{tabular}{@{}l l r r l r r r r r@{}}
\toprule
 & \multicolumn{4}{c}{\textbf{Setup}} & \multicolumn{5}{c}{\textbf{Cost}} \\
\cmidrule(lr){2-5}\cmidrule(l){6-10}
\textbf{Experiment} & \textbf{Attacker model} & $k$ & \textbf{Scen.} & \textbf{GPUs} & \textbf{Calls} & \textbf{Wall} & \textbf{GPU-h} & s/c & tok \\
\midrule
\multicolumn{10}{@{}l}{\textit{AgentDojo v1.2.2, target Qwen3.6-35B served with vLLM, thinking off}}\\
PIMiner & Qwen3.6-35B, thinking on & 100 & 949 & 3 B, 2 A & 39.7 & 51$^{\P}$ & $\approx$220$^{\P}$ & 29$^{\P}$ & $\approx$11.5 \\
CtH & Released Qwen3-4B & 100 & 949 & 3 B, 2 A & 94.9 & 15 & $\approx$63$^{\|}$ & 2.1 & $\approx$1.5 \\
AutoDojo & Qwen3.6-35B, thinking on & 99+5$^{\ast}$ & 389 & 2 B$^{\ast}$ & 48.2 & $\approx$28 & $\approx$51 & 3.8 & 2.3 \\
\rowcolor{tabalt}
\sysL{} & Qwen3.6-35B & 133 & 946 & 3 B, 2 A & 91.0 & n/a & $\approx$29 & 1.1 & 0.5 \\
\rowcolor{tabalt}
\sysF{} & Qwen3.6-35B & 529 & 946 & 3 B, 2 A & 305.4 & 23 & $\approx$96 & 1.1 & 0.5 \\
\rowcolor{tabalt}
\sysSE{} & Qwen3.6-35B & 601 & 948 & 3 B, 4 A & 229.1 & 33 & $\approx$145 & 2.3 & 0.5 \\
\midrule
\multicolumn{10}{@{}l}{\textit{AgentDojo, target Haiku~5.5 through its API with default settings, 97 sampled scenarios (one injection task per user task)}}\\
PIMiner & As above & 50--200\rlap{\smash{$^{\ddagger}$}} & 97 & 2 B & 6.1 & 53 & $\approx$105 & 62 & n/a \\
CtH & As above & 100 & 97 & 2 A & 9.7 & 1.9 & $\approx$1.5 & 0.6 & n/a \\
AutoDojo & Qwen3.6-35B, thinking off & 30 & 57 & 1 B & 1.6 & n/a & $\approx$1.2 & 2.7 & n/a \\
\rowcolor{tabalt}
\sysL{} & As above & 133 & 97 & 2 B & 11.4 & 2.0 & $\approx$4 & 1.1 & n/a \\
\rowcolor{tabalt}
\sysF{} & As above & 529 & 97 & 2 B & 44.5 & 5.9 & $\approx$12 & 0.9 & n/a \\
\rowcolor{tabalt}
\sysSE{} & As above & 601 & 97 & 3 B, 2 A & 54.8 & 7.8 & $\approx$19 & 1.3 & n/a \\
\midrule
\multicolumn{10}{@{}l}{\textit{IPI Arena, 41 scenarios, target Qwen3.6-35B served with vLLM, thinking off$^{\S}$}}\\
PIMiner & As on AgentDojo & 100 & 41 & 2 B & 2.7 & 6 & $\approx$12 & 16 & 8.5 \\
CtH & As on AgentDojo & 100 & 41 & 2 B & 4.1 & 0.9 & $\approx$2 & 1.7 & 1.2 \\
\rowcolor{tabalt}
\sysL{} & As on AgentDojo & 133 & 41 & 2 B & 3.5 & n/a & $\approx$2 & 2.1 & 1.5 \\
\rowcolor{tabalt}
\sysF{} & As on AgentDojo & 529 & 41 & 2 B & 11.4 & 7.7 & $\approx$6.5 & 2.1 & 1.5 \\
\rowcolor{tabalt}
\sysSE{} & As on AgentDojo & 601 & 41 & 2 B & 18.5 & 8.8 & $\approx$11 & 2.1 & 1.5 \\
\bottomrule
\end{tabular}
\\[1pt]
{\scriptsize The static seed (\cref{sec:method:loop}) is attempt 1 of the \sysL{} run on each target. On Qwen3.6-35B the \sysL{} and \sysF{} run lacks three AgentDojo scenarios and \sysSE{} one. CtH draws all its samples up front, and PIMiner's library is frozen. $^{\ast}$Includes five replays per scenario on 2 A. $^{\ddagger}$50 for each of up to 4 injection points. $^{\S}$The same servers run IPI Arena's judge and tool simulator and score all criteria, PIMiner's attacker has a 131K-token context, and CtH uses its IPI Arena prompt. $^{\P}$One run of 27.0 thousand calls. $^{\|}$12\% repeated five slow units, left out of s/c. Costs exclude one-time training and CtH's sample generation, \sysL{} costs on Qwen3.6-35B are prorated from \sysF{}, GPU-h and s/c ignore the GPU model and cover only the attacker side for Haiku~5.5, without API fees, and the two $\approx$ tok values are estimates.}
\end{table*}

\paragraph*{Implementation}
The engine core is about 2,100 lines of Python with no benchmark dependency,
and each data point writes a JSON trace of every attempt. Each generator
response is parsed between explicit delimiters and sanitized by a routine
written for AgentDojo's YAML environment files that runs unchanged on every
benchmark. A line that fails a YAML round-trip is discarded like one that
cannot be parsed (\cref{sec:method:loop}). The generator is an open-weight
model that we serve locally, so no provider's safeguards are involved. The
taxonomy descriptions it sees are lightly neutralized with their meaning
preserved, and goal identifiers are masked out of judge feedback
(\cref{sec:method:adaptive}).

\paragraph*{Compute and reproducibility}
\sys{} uses generative models as the object of study (targets) and as part of
the method (generator, judge). All open models ran on three NVIDIA RTX PRO 6000
Blackwell (98\,GB) and up to four RTX A6000 (48\,GB) GPUs across four servers,
and the hosted target was queried through its public API. The \sysL{} runs on
our three settings issue about 106,000 target calls in about 35 GPU-hours. To
keep this footprint low we use one mid-sized open model (35B parameters, about
3B active, in FP8) for every attacker role and stop each data point at its
first success. Hosted targets can change or be retired, so we record versions and
dates and release every generated injection for re-scoring, with the engine, the adapters, and the configurations.

\section{The Taxonomy File}
\label{app:taxonomy}
Our taxonomy file follows the published \tax{}
listing~\cite{damianakis2026piatlas}, which has 131 sub-techniques, and adds one
entry for non-semantic sequence (F3.T6), which has none, so that it is still
swept. That gives 132. \cref{tab:piatlas} lists its techniques. \sys{} reads the
taxonomy from a file and can use another one that meets four requirements. It
has three levels of units, each with an identifier and a name. Each
sub-technique has a definition and optionally one example, which is all the
generator learns about it (\cref{sec:method:loop}). Every unit describes how the
injected text is built, so that the generator can realize it while the goal
stays fixed, which a taxonomy by attacker goal
does not allow. The units do not overlap, so that the credit for each success is
unambiguous.  Our order front-loads three of the five
structural-spoofing techniques, and obfuscation is followed only by the four
techniques that the order does not yet rank, two of which are also
obfuscation.

\section{Post-hoc Technique Labeling}
\label{app:labeling}
No attacker's own labels enter our coverage numbers. We label the first
successful injection of every broken scenario in every run of
\cref{tab:runsettings}, \sys{}'s included, 3,171 broken scenarios and 2,485
distinct texts in all, by one procedure that sees only the text, not the
attacker, the scenario, or the sub-technique the text was generated from. It
follows the labeling of \corp{}~\cite{damianakis2026piatlas} and extends it to
sub-techniques. When a scenario has multiple texts, its labels are the union
over those texts.

\paragraph*{Annotators and consensus}
Each text is labeled by three independent calls to Claude Opus~5.5 at high
reasoning effort, one text per call, in a fresh context without tools. One
fixed prompt lists every unit of the taxonomy file with its definition and asks
for every sub-technique whose construction is present in the text, each with a
verbatim quote as evidence. A unit counts for a text when at least two
annotators give it, voted separately at each level, and the Haiku~5.5 coverage
of \cref{tab:results} requires all three.
A technique can thus count while none of its sub-techniques does, as for
AutoDojo against Haiku~5.5. Its four successes are one text, the bare injection goal,
which all three annotators label a plain command but assign to two different
sub-techniques. The mean pairwise Jaccard agreement is
0.84, 0.90, and 0.96 at the sub-technique, technique, and class levels, and
97\% of the evidence quotes appear verbatim in the text.

\paragraph*{Validation and exclusions}
We fixed the annotator settings on 64 calibration texts from every attacker.
Pilot calls at three effort levels disagreed on 120 labels, which one author
judged without seeing which call had given them. Against the resulting
reference, a single annotator call reaches a precision of 0.89 and a recall of
0.87 at the sub-technique level, and an F1 of 0.89 and 0.95 at the technique
and class levels.  The provider's safety filter stopped 384 of the 7,455 calls
(5.2\%), which we did not retry. The 119 texts (4.8\%) left with fewer than two
annotators are excluded, which removes 123 of the 3,171 broken scenarios, 62 of
PIMiner's and 41 of CtH's on AgentDojo against Qwen3.6-35B, 10\% and 5\% of
their broken scenarios there, and at most three per run elsewhere.
The counts in \cref{tab:results,tab:covcheck} are of labeled scenarios.

\paragraph*{Intended and detected labels}
Of the 132 labeled scenarios that the sweep broke on AgentDojo against
Qwen3.6-35B, 97 (73\%) carry the sub-technique the line was generated from, 103
(78\%) its technique, and 112 (85\%) its class. A sweep injection receives 2.1
sub-technique labels on average, against 5 for the AgentDojo seed and 5 to 6 for
the baselines.

\section{Prompt Structure}
\label{app:prompts}
The exact prompts ship with the released engine and follow
\cref{sec:method:loop,sec:method:adaptive}. The exploit prompt lists the last 10
failed rewrites of a line and the judge feedback on them, or the last 24 in the
AgentDojo \sysSE{} run on Qwen3.6-35B. The judge sees the user's request, the
agent's tool calls with their arguments, and its visible replies, each cut to a
fixed length, and context blending uses the preceding transcript. Neither reads
the system prompt or hidden reasoning. Qwen3.6-35B runs with thinking off, and
for Haiku~5.5 our adapter keeps only the visible text and the tool calls of each
reply.

\section{Cost and Equal-Cost Budgets}
\label{app:attemptcost}
The target side of an attempt is the same for every attacker, one agent run on
the scenario. The attackers differ in the work they do to produce a candidate,
in how many attempts run at once, and in what they pay before the first
attempt. A PIMiner attempt is an agentic iteration over a long strategy library
and the previous trajectory, and iterations on a scenario run in sequence,
whereas \sys{} adds one short generator call to each target run and runs
scenarios in parallel. In our Haiku~5.5 runs two GPUs produced about 100 \sys{}
attempts per minute and about 2 PIMiner attempts per minute.
\cref{tab:runsettings} lists the cost of every experiment. On IPI Arena the
generator returned no usable line for 24\% of the sweep attempts of \sysF{},
because it declined some of the goals, so an unbroken scenario received 378
target runs on average, not 529.

\paragraph*{Work per attempt}
GPU-hours count the hours a GPU was assigned to a run, including the end of a
run, when the GPUs are mostly idle, and they mix two GPU models. For the
equal-cost budgets we therefore count work, the tokens the model servers
generated per target call (tok in \cref{tab:runsettings}), read from the
servers' logs. A busy server generated 600 to 1,035 tokens per second in every
run, so we convert at 800 tokens per GPU-second. On AgentDojo that gives 0.6
GPU-seconds per attempt for \sys{}, 1.8 for CtH, 2.9 for AutoDojo, and 14.4 for
PIMiner, and on IPI Arena 1.8 for \sys{}, 1.6 for CtH, and 10.6 for PIMiner.
\sysSE{} runs the engine of \sysF{} and shares its value. Two values
are estimates. CtH's on AgentDojo is measured on three of the four suites, and
PIMiner's is its logged attacker output plus a target run like CtH's, because
the server logs of that run were overwritten.

\paragraph*{Training}
\sys{} and AutoDojo train nothing. PIMiner's library mining took 633 target
calls and about one hour on two GPUs in our port. One CtH curriculum stage
against Qwen3.6-35B with the released configuration made 28,800 target calls in
13 hours on two GPUs, or 26 GPU-hours. The released attacker went through three stages, against GPT-5 nano,
GPT-5.6 Luna, and GPT-5.6 Terra, about 90,000 target calls with the tests
between stages and about 78 GPU-hours had the targets been local. The CtH paper reports no training cost, and its authors lacked the
budget for a stage against the next, more robust model. \cref{fig:efficiency}
spreads the 78 GPU-hours over the 99,000 attempts CtH made against Qwen3.6-35B
on the two benchmarks, which adds 2.8 GPU-seconds per attempt to the 1.8 of run
time and gives the 4.7 of the figure.

\paragraph*{Equal-cost budgets}
Attackers stop
a scenario at different points, so \cref{tab:budgets} equalizes the whole run.
Every attacker gets what \sysF{} spent on all of AgentDojo, 147 million
generated tokens or 51 GPU-hours. PIMiner uses it at about 32 attempts, as its
attempts grow from 2,500 tokens to an average of 11,500, and CtH at 41 with its
training share, while its 100 attempts fit with none charged. AutoDojo's
optimization generated 107 million tokens on three suites, about three times
the share of \sysF{} there, and \sysSE{} used three quarters of it in
601 attempts. On IPI Arena \sysF{} made 11,401 target runs, which gives PIMiner
about 65 attempts, CtH at most 115, and \sysSE{} about 350
(\cref{tab:budgets}). PIMiner's ASR@$k$ is flat after attempt 44, and CtH ran
100 attempts, a lower bound at equal cost. We set no equal-cost budgets for
Haiku~5.5, whose costs leave out API fees.

\section{Circularity Checks for Coverage}
\label{app:covcheck}
\begin{table}[t]
\centering
\caption{Circularity checks on Qwen3.6-35B at equal sample size $k$. $\mathrm{Cov}$
counts classes over all successes, $\mathrm{Cov}@k$ distinct classes and techniques
among $k$, $\bar d$ and VS are embedding diversity. Best per block in \textbf{bold}.}
\label{tab:covcheck}
\footnotesize
\renewcommand{\arraystretch}{1.0}
\setlength{\tabcolsep}{3.2pt}
\begin{tabular}{@{}l c cc cc@{}}
\toprule
 & $\mathrm{Cov}$ & \multicolumn{2}{c}{$\mathrm{Cov}@k$} & \multicolumn{2}{c}{Embedding} \\
\cmidrule(lr){2-2}\cmidrule(lr){3-4}\cmidrule(l){5-6}
\textbf{Attacker} & class & class & tech. & $\bar d$ & VS \\
\midrule
\multicolumn{6}{@{}l}{\textit{AgentDojo, $k=98$}}\\
Static seeds         & 3 & 3.00 & 5.0 & 0.00 & 1.0 \\
PIMiner              & 6 & 5.18 & 20.4 & \underline{0.45} & \textbf{9.7} \\
CtH                  & 6 & 5.12 & 16.5 & 0.32 & 5.9 \\
AutoDojo             & 5 & 5.00 & 18.9 & 0.26 & 3.4 \\
\rowcolor{tabalt}
\sysL{} (ours)       & 6 & \underline{5.89} & \underline{21.3} & 0.43 & 6.7 \\
\rowcolor{tabalt}
\sysF{} (ours)       & 6 & \textbf{5.97} & \textbf{22.9} & \textbf{0.47} & \textbf{9.7} \\
\rowcolor{tabalt}
\sysSE{} (ours)      & 5 & 4.92 & 14.9 & 0.41 & \underline{8.0} \\
\multicolumn{6}{@{}l}{\textit{IPI Arena, $k=8$}}\\
PIMiner              & \textbf{5} & \textbf{5.00} & \textbf{16.9} & \underline{0.57} & 5.12 \\
CtH                  & \textbf{5} & \underline{4.91} & \underline{13.5} & 0.52 & 4.52 \\
\rowcolor{tabalt}
\sysL{} (ours)       & 3 & 3.00 & 6.0 & \textbf{0.61} & \textbf{5.46} \\
\rowcolor{tabalt}
\sysF{} (ours)       & \underline{4} & 3.67 & 8.8 & \textbf{0.61} & \underline{5.39} \\
\rowcolor{tabalt}
\sysSE{} (ours)      & 3 & 2.80 & 7.4 & 0.51 & 4.51 \\
\bottomrule
\end{tabular}
\end{table}

\sys{} generates its injections from \tax{} and coverage is counted in \tax{},
so high coverage could repeat how the injections were produced and labeled
rather than show real variety (\cref{sec:setup:metrics}). \cref{tab:covcheck}
reports two checks, coverage at an equal sample size and embedding diversity,
which uses no taxonomy.

\paragraph*{Sample and labels}
For an attacker $a$ on one target and benchmark, let $S_a$ hold the first
successful injection of each broken scenario, labeled as in \cref{app:labeling}.
Coverage and the Vendi score grow with the number of injections, so attackers
are compared at an equal sample size $k=\min_a|S_a|$. Each value is the mean
over all subsets of $S_a$ of size $k$, or over 5{,}000 random ones when there
are more, and $\mathrm{Cov}@k$ is the mean number of distinct units among the
labels of $k$ injections.

\paragraph*{Embedding diversity}
In each injection we replace the goal and the entities it names (quoted
strings, addresses, account numbers, identifiers, and names) with placeholders,
so that two injections differ in how they deliver the goal and not in what they
ask for. An encoder maps each text to a unit vector $x_i$, and an injection
with several texts is the mean of their vectors. We use the encoder of
\tax{}~\cite{damianakis2026piatlas}, F2LLM-v2-14B instructed to represent the
rhetorical strategy of a text rather than its topic, and repeat the analysis
with Qwen3-Embedding-8B without an instruction. For $k$ injections with
$K_{ij}=x_i^{\top}x_j$ we report the mean pairwise cosine distance
$\bar d=\binom{k}{2}^{-1}\sum_{i<j}(1-K_{ij})$ and the Vendi score
$\mathrm{VS}=\exp(-\sum_i\lambda_i\ln\lambda_i)$~\cite{friedman2023vendi}, where
the $\lambda_i$ are the eigenvalues of $K/k$. $\mathrm{VS}$ is the effective
number of distinct injections, from 1 for identical texts to $k$ for
orthogonal ones, so it is comparable only within one experiment, whereas
$\bar d$ does not depend on $k$. Surface edits can inflate embedding
distances~\cite{wen2025rl}, so they serve only as an independent check.

\paragraph*{Results}
On AgentDojo against Qwen3.6-35B, $k=98$ is the number of scenarios that the
static seed breaks. With
Qwen3-Embedding-8B, PIMiner leads (0.53 and 13.9), ahead of \sysSE{},
CtH, and \sysF{} (0.48 to 0.50, 10.6 to 11.8). Both encoders place AutoDojo last
among the adaptive attackers. On the scenarios that the static seed does not
break ($k=132$), \sysF{} and \sysL{} cover 26.0 and 25.0 techniques, against
20.8 for PIMiner, 20.5 for AutoDojo, and 17.1 for CtH, and both lead on
distance and Vendi score with either encoder (0.52 and 14.9 for \sysF{} against
0.44 and 10.1 for PIMiner with the primary one). \cref{fig:embmap} shows this
sample, laid out by multidimensional scaling of the cosine distances within and
then between attackers ($r=0.71$ between drawn and true distances), with a
larger mark for several identical texts.
The static seed is one mark of fixed
size, because its 98 successes share one template and differ only in the
masked goal.

On IPI Arena ($k=8$) an injection of PIMiner carries 5.2 technique labels and one of \sysF{} 1.6, and AgentDojo cut to
eight successes gives 14.3 techniques for PIMiner against 10.1. The measures that do not grow with the number of labels
show no deficit. Two successes of \sysF{} share
fewer techniques than two of PIMiner or CtH ($\bar J$ of 0.90 against 0.75 and
0.76), and its injections lie farther apart than those of PIMiner with both
encoders (0.61 against 0.57, and 0.62 against 0.59), though CtH is first with
Qwen3-Embedding-8B (0.67). The sweep there also tried fewer sub-techniques,
because the generator declined some goals (\cref{app:attemptcost}).

Against Haiku~5.5, as in \cref{fig:embmap}, we measure embedding distance only
on the scenarios that the static seed does not break. Only \sysF{} keeps two
successes on them, with a mean embedding distance of 0.55 (0.64 with
Qwen3-Embedding-8B), and no other attacker keeps more than one.

\section{Search Strategies}
\label{app:strategies}
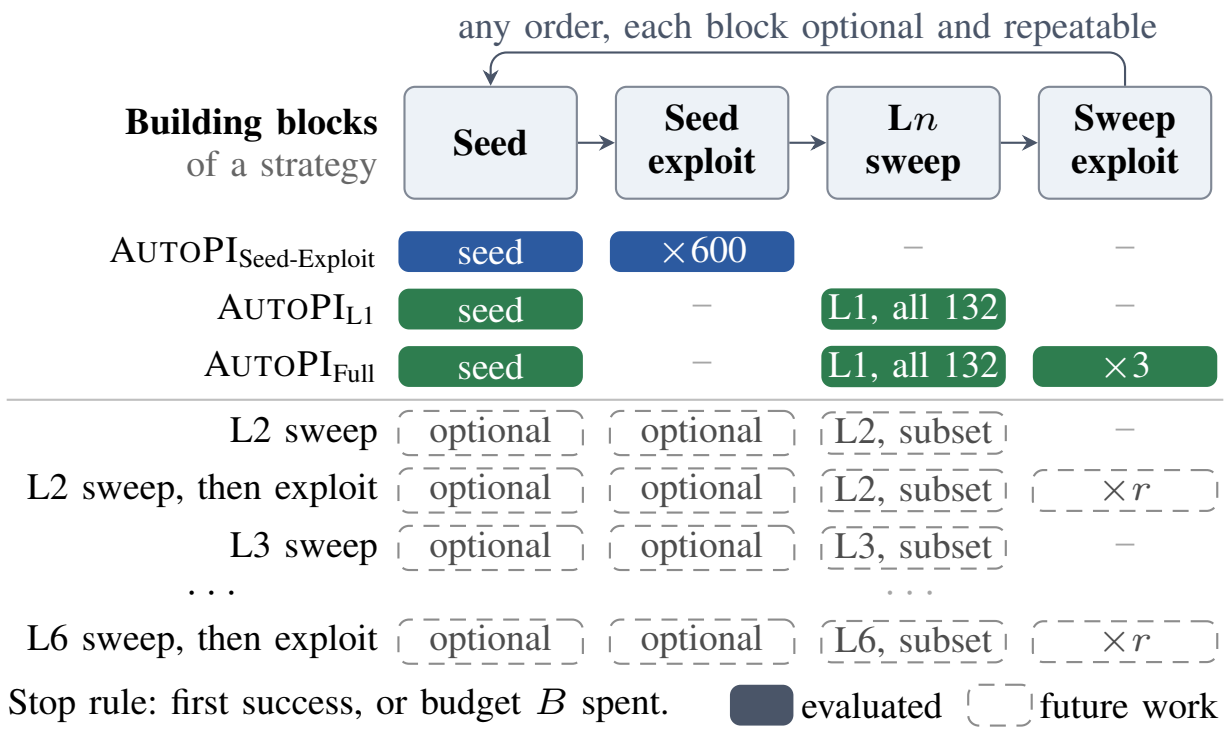
\begin{figure}[t]
\centering
\begin{tikzpicture}[x=0.985cm, y=1cm, font=\scriptsize,
  blk/.style={draw=sysgray!70, fill=tabhdr, rounded corners=2pt, minimum width=1.14cm, minimum height=0.74cm,
              align=center, inner sep=1pt, font=\scriptsize\bfseries, text=black},
  done/.style={fill=#1, text=white, rounded corners=2pt, minimum width=1.22cm, minimum height=0.27cm, inner sep=1pt},
  todo/.style={draw=black!45, densely dashed, text=black!70, rounded corners=2pt, minimum width=1.22cm,
               minimum height=0.27cm, inner sep=1pt},
  none/.style={text=black!35},
  lab/.style={anchor=east, inner sep=0pt}]
  \node[blk] (b1) at (3.60,0.40) {Seed};
  \node[blk] (b2) at (5.02,0.40) {Seed\\exploit};
  \node[blk] (b3) at (6.44,0.40) {L$n$\\sweep};
  \node[blk] (b4) at (7.86,0.40) {Sweep\\exploit};
  \draw[-{Stealth[length=3pt,width=3pt]}, sysgray] (b1) -- (b2);
  \draw[-{Stealth[length=3pt,width=3pt]}, sysgray] (b2) -- (b3);
  \draw[-{Stealth[length=3pt,width=3pt]}, sysgray] (b3) -- (b4);
  \draw[-{Stealth[length=3pt,width=3pt]}, sysgray, rounded corners=3pt]
    (b4.north) -- ++(0,0.22) -| (b1.north);
  \node[anchor=south, text=sysgray, inner sep=1pt] at (5.73,1.00) {any order, each block optional and repeatable};
  \node[lab, font=\scriptsize\bfseries] at (2.85,0.52) {Building blocks};
  \node[lab, text=black!60] at (2.85,0.24) {of a strategy};
  \node[lab] at (2.85,-0.32) {\sysSE{}};
  \node[done=sysblue] at (3.60,-0.32) {seed};
  \node[done=sysblue] at (5.02,-0.32) {$\times$600};
  \node[none] at (6.44,-0.32) {--}; \node[none] at (7.86,-0.32) {--};
  \node[lab] at (2.85,-0.70) {\sysL{}};
  \node[done=sysgreen] at (3.60,-0.70) {seed};
  \node[none] at (5.02,-0.70) {--};
  \node[done=sysgreen] at (6.44,-0.70) {L1, all 132};
  \node[none] at (7.86,-0.70) {--};
  \node[lab] at (2.85,-1.08) {\sysF{}};
  \node[done=sysgreen] at (3.60,-1.08) {seed};
  \node[none] at (5.02,-1.08) {--};
  \node[done=sysgreen] at (6.44,-1.08) {L1, all 132};
  \node[done=sysgreen] at (7.86,-1.08) {$\times$3};
  \draw[black!25, line width=0.4pt] (0.35,-1.30) -- (8.52,-1.30);
  \node[lab] at (2.85,-1.52) {L2 sweep};
  \node[todo] at (3.60,-1.52) {optional};
  \node[todo] at (5.02,-1.52) {optional};
  \node[todo] at (6.44,-1.52) {L2, subset};
  \node[none] at (7.86,-1.52) {--};
  \node[lab] at (2.85,-1.90) {L2 sweep, then exploit};
  \node[todo] at (3.60,-1.90) {optional};
  \node[todo] at (5.02,-1.90) {optional};
  \node[todo] at (6.44,-1.90) {L2, subset};
  \node[todo] at (7.86,-1.90) {$\times r$};
  \node[lab] at (2.85,-2.28) {L3 sweep};
  \node[todo] at (3.60,-2.28) {optional};
  \node[todo] at (5.02,-2.28) {optional};
  \node[todo] at (6.44,-2.28) {L3, subset};
  \node[none] at (7.86,-2.28) {--};
  \node[inner sep=0pt] at (1.75,-2.58) {\raisebox{-1pt}{$\cdots$}};
  \node[none] at (6.44,-2.58) {$\cdots$};
  \node[lab] at (2.85,-2.90) {L6 sweep, then exploit};
  \node[todo] at (3.60,-2.90) {optional};
  \node[todo] at (5.02,-2.90) {optional};
  \node[todo] at (6.44,-2.90) {L6, subset};
  \node[todo] at (7.86,-2.90) {$\times r$};
  \node[anchor=west, inner sep=0pt] at (0.35,-3.32) {Stop rule: first success, or budget $B$ spent.};
  \node[done=sysgray, minimum width=0.42cm] at (5.42,-3.32) {};
  \node[anchor=west, inner sep=0pt] at (5.68,-3.32) {evaluated};
  \node[todo, minimum width=0.42cm] at (7.02,-3.32) {};
  \node[anchor=west, inner sep=0pt] at (7.28,-3.32) {future work};
\end{tikzpicture}
\caption{Search strategies built from four blocks. L$n$ stacks $n$
sub-techniques per injection, and $\times r$ counts exploits per failed line.}
\label{fig:strategies}
\end{figure}
A strategy combines the four blocks of \cref{fig:strategies}. The \emph{seed} tries the fixed seed, a \emph{seed exploit}
$\times r_0$ tries $r_0$ rewrites of it, an \emph{L$n$ sweep} of $T$ tries one
line for each set of $n$ sub-techniques of a subset $T$ of the taxonomy, in a
given order, and a \emph{sweep exploit} $\times r$ tries $r$ variants of
each failed sweep line. A \emph{search strategy}
$\sigma=(\phi_1,\dots,\phi_m,B,\rho)$ is a finite list of blocks with their parameters, in
which a block can be absent or repeated, plus an attempt budget $B$ per data point (scenario) and
a stop rule $\rho$, either the first success or the end of the budget.
\cref{alg:loop} runs $\sigma$ on one data point. \sysSE{} is the seed
and a seed exploit $\times 600$ ($B=601$). \sysF{} is the seed, an L1 sweep
of all 132 sub-techniques in the priority order of \cref{sec:method:loop}, and
a sweep exploit $\times 3$ ($B=1+132\cdot 4=529$), and \sysL{} stops
before the sweep exploit ($B=133$). Candidates grow quickly with $n$. An L1
sweep has 132, an L2 sweep 8,646, an L3 sweep 374,660, and an L6 sweep over six
billion, so higher levels sweep only the subset that the budget allows, for
example the sub-techniques most promising in historical data. Which strategy is
best depends on the objective (coverage, ASR, or efficiency) and on the target
and benchmark, and finding it is future work.

\end{document}